\documentclass[aps,prb,twocolumn,superscriptaddress,showpacs]{revtex4}

\usepackage{times}
\usepackage{epsfig}
\usepackage{amsmath}
\usepackage{amssymb}

\begin{document}

\title{Exotic resonant level models in non-Abelian quantum Hall states
coupled to quantum dots}

\author{Gregory A. Fiete}
\affiliation{Department of Physics, The University of Texas at Austin, Austin, Texas 78712, USA}
\author{Waheb Bishara}
\affiliation{Department of Physics, California Institute of Technology, MC 114-36, Pasadena, California 91125, USA}
\author{Chetan Nayak}
\affiliation{Microsoft Research, Station Q, CNSI Building, University of California, Santa Barbara, California 93106, USA}
\affiliation{Department of Physics, University of California, Santa Barbara, California, 93106, USA}

\date{\today}

\begin{abstract}

In this paper we study the coupling between a quantum dot and the edge of a
non-Abelian fractional quantum Hall state.   We assume the dot is small enough that its level spacing is large compared to both the temperature and the
coupling to the spatially proximate bulk non-Abelian fractional quantum Hall state.  We focus on the physics of level degeneracy with electron number on the dot.  The physics of such a resonant level is governed by a $k$-channel Kondo model when the quantum Hall state is a Read-Rezayi state at filling fraction $\nu=2+k/(k+2)$ or its particle-hole conjugate at $\nu=2+2/(k+2)$. The $k$-channel Kondo model is channel symmetric even without fine tuning any couplings in the former state; in the latter, it is generically channel asymmetric.   The two limits exhibit non-Fermi liquid and Fermi liquid properties, respectively, and therefore may be distinguished.   By exploiting the mapping between the resonant level model and the multichannel Kondo model, we discuss the thermodynamic and transport properties of the system.   In the special case of $k=2$, our results provide a novel venue to distinguish between the Pfaffian and anti-Pfaffian states at filling fraction $\nu=5/2$.  We present
numerical estimates for realizing this scenario in experiment.   

 \end{abstract}



\maketitle

\section{Introduction}
One of the most outstanding problems in the fractional quantum Hall effect is the unambiguous experimental identification of a state with non-Abelian quasi-particle braiding statistics.\cite{Nayak:rmp08}  To date, there is no physical system that has been conclusively demonstrated to have this property. Certain fractional quantum Hall states, most notably $\nu=5/2$ and $\nu=12/5$, are believed to be exceptionally promising candidates for possessing non-Abelian quasi-particles and their experimental verification would constitute a major milestone in physics.  The
evidence that the $\nu=5/2$ quantum Hall state
is non-Abelian comes from a combination of analytical theory;
\cite{Moore:npb91,Greiter92,Nayak96c}
numerical study of small systems;
\cite{Morf98,Rezayi00,Wang09,Bonderson09,Feiguin:prb09,Wojs:prb09}
and, most importantly, experiment itself.\cite{Dolev:nat08,Radu08,Willett:pnas09}
In the case of the $\nu=12/5$ state, \cite{Read:prb99,Bonderson08}
conjectures of non-Abelian-ness
are based solely on numerics.\cite{Rezayi:prb09,Bonderson09,Wojs:prb09}

Non-Abelian quantum Hall states were first constructed with the use of conformal field theory (CFT).\cite{CFT}   Moore and Read\cite{Moore:npb91} argued that candidate wavefunctions could be computed as certain correlation functions (conformal blocks) of a CFT.  By requiring that the wavefunctions have certain physical properties, the set of allowed CFTs turn out to be highly constrained.  The simplest CFT that satisfies the constraints and possess non-Abelian quasi-particles is the $\mathbb{Z}_2$ Ising CFT which leads to a wavefunction known as the Moore-Read (MR) ``Pfaffian" state,
so named because the wavefunction contains a Pfaffian factor. Numerics
\cite{Morf98,Rezayi00,Feiguin:prl08,Wojs:prb09} supports this wavefunction as a
candidate description of the $\nu=5/2$ plateau,\cite{Willett:prl87,Pan:prl99} which has the largest gap among observed
plateaus in the second Landau level.\cite{Xia04,Choi07,Dean:prl08}

The hypothesis that the $\nu=5/2$ state is non-Abelian
has, until recently, been based on the Moore-Read (MR)
wavefunction \cite{Moore:npb91} and quasiparticle excitations
above it.\cite{Nayak96c} However, it was recently
realized that there is an alternate non-Abelian candidate
for $\nu=5/2$, the {\it anti-Pfaffian} state \cite{Levin:prl07,Lee:prl07}
which, in the limit of vanishing Landau-level mixing,
is exactly degenerate in energy with the MR Pfaffian
state (although both might, in principle, be higher in energy than some other state).
Exact diagonalization studies have generally neglected Landau-level mixing
and have, therefore, not distinguished between these two states.
Early numerics on the sphere \cite{Morf98} failed to look to for the anti-Pfaffian
state, which occurs at a different shift. On the torus, \cite{Rezayi00}
finite-size effects, which cause mixing between these two states and a
consequent energy splitting, were too large. However, more recent
numerics on the sphere \cite{Feiguin:prl08} and the torus \cite{Peterson:prb08}
make it clear that, in the absence of Landau-level mixing,
the anti-Pfaffian state is as good a candidate
as the MR Pfaffian state with a ground state energy
which converges with that of the MR Pfaffian in the thermodynamic limit.

Recent work on the effects of Landau-level mixing indicate that the anti-Pfaffian
state is, in fact, lower in energy for realistic values of the magnetic field.
\cite{Bishara09,Wang09} However, it is difficult to experimentally
distinguish the anti-Pfaffian state from the MR Pfaffian state.
The smallest quasiparticle charge is $e/4$ in both states,
so shot noise measurements \cite{Dolev:nat08} cannot distinguish
between them. The two states' non-Abelian braiding properties are very
similar. Thus, interferometry experiments, \cite{Bonderson:prl06,Bonderson_2:prl06,Stern:prl06,Feldman:prl06,Sarma:prl05,Fidkowski:cm07,Bishara:prb08,Bishara09} such as
those of Willett et al. \cite{Willett:pnas09} are also unlikely
to distinguish between these two states. 
It is possible, in principle, to distinguish them by their
$I(V,T)$ curves, but these can be modified by edge reconstruction,
so the distinction between the $I(V,T)$ curves of the two states might be blurred.
The two states can also be distinguished by the signs of their thermal Hall
conductances, but only if the larger contribution of the filled lowest Landau level
(of both spins) can be separated. Therefore, there is a need for experiments
capable of distinguishing these two candidate descriptions of the
$\nu=5/2$ state. In this paper, we show how differences in the gapless edge modes of
the anti-Pfaffian and MR Pfaffian states lead to different quantum impurity
(resonant level) models and transport properties when
these states are coupled to quantum dots.\cite{Fiete:prl08}  We also briefly discuss the expectations for other (numerically less competitive) quantum Hall states at $\nu=5/2$, such as the SU(2)$_2$ NAF,\cite{Wen:prl91,Blok:npb92} $K=8$ strong pairing,\cite{Halperin83} and (3,3,1) state.\cite{Halperin83}

The MR Pfaffian state was generalized by
Read and Rezayi \cite{Read:prb99} (RR), who constructed
wavefunctions using the $\mathbb{Z}_k$ parafermion CFTs
(the case $k=2$ is the MR Pfaffian wavefunction).  These wavefunctions are
candidate states for filling fraction $\nu=2+k/(k+2)$ and they all
posses non-Abelian quasi-particle braiding statistics for $k\geq2$.
The particle-hole conjugates of the RR states, which we will call
the `anti-RR states', were constructed recently; \cite{Bishara:prbR08}
the $k=3$ anti-RR state is a candidate description
of the $\nu=12/5$ plateau. Another class of non-Abelian candidate
states follow from Bonderson and Slingerland's \cite{Bonderson08}
(BS) hierarchy construction built on the $\nu=5/2$ state. At $\nu=12/5$, the
$k=3$ anti-RR state and one of the BS states are distinct non-Abelian 
candidates. Our setup can be applied to distinguish these two states.

\begin{figure}[t]
\includegraphics[width=.65\linewidth,clip=]{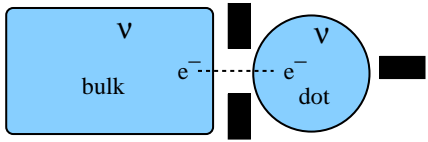}
 \caption{\label{fig:schematic} (color online) Schematic of our model.  Gates are shown in black. They may be used to form a point contact to pinch off the dot from the rest of the quantum Hall bulk.  The gate on the right of the figure may be used to shift the energy levels of the dot by changing its area $S$.  The bulk is assumed to be at filling fraction $\nu=2+k/(k+2)$ or $\nu=2+2/(k+2)$. The white region between the dot
and the bulk is assumed to be at $\nu=2$.
The charge on the dot may be measured capacitively.
\cite{Berman:prl99,Duncan:apl99}}
 \end{figure}

A subset of the results presented here was succinctly given in Ref.~[\onlinecite{Fiete:prl08}].  The basic scenario we discuss is shown in Fig.\ref{fig:schematic}.
 Our main result is that electron tunneling between a level-$k$ RR
 state and a quantum dot at its degeneracy point, $E(N+1)=E(N)$, can be mapped to a {\em stable} $k$-channel,  channel symmetric,  Kondo model (where $k\geq2$ is given by the filling fraction $\nu=2+k/(k+2)$, assumed to be described by a RR state).  Here $E(N)$ is the energy of the quantum dot with $N$ electrons on it.  Since the non-Fermi liquid physics of multi-channel Kondo models with channel isotropy is usually unstable to small anisotropies in the coupling to different channels, this physics is very difficult to realize in experiment.  A central result of this work is that the non-Abelian nature of the RR fractional quantum Hall states {\em automatically} provides channel isotropy in the emergent multi-channel Kondo model.  In this sense, the non-Abelian fractional quantum Hall state provides  ``topological protection" to the non-Fermi liquid physics of the multi-channel Kondo model.   In this way, we have uncovered a
scenario in which a global property, the topological order of the RR states,
protects fragile {\em local} quantum impurity physics.  Our analysis is based
on properties of the gapless edge excitations required by the topological order of the RR states.

We also consider the situation in which two bulk quantum Hall droplets
are separated by a quantum dot (i.e. another bulk droplet on the right
side of the dot in Fig.\ref{fig:schematic}, as shown in Fig.\ref{fig:transport}). This setup is relevant to a
two point-contact interferometer \cite{Chamon:prb97,Fradkin98,Bonderson_2:prl06,Bonderson:prl06,
Stern:prl06,Willett:pnas09} when the backscattering at each point contact
is large, so that it is in the Coulomb blockade regime.\cite{Rosenow:prl07}  One may naively think that this scenario would ``double" the number of ``conduction" electron channels and lead to a $2k$-channel Kondo model.  However, we show explicitly that this is not the case and that its true fixed point (for equal electron tunneling to both ``bulk" states) is more subtle.

This paper is organized as follows.  In Sec.~\ref{sec:edge_theory} we introduce the essential elements of the edge theory of Abelian and non-Abelian fractional quantum Hall edge states, including the particle-hole conjugate states.  In Sec.~\ref{sec:CFT_Kondo} we review the Emery-Kivelson solution to the 2-channel Kondo model and the more general Affleck-Ludwig CFT solution to the $k$-channel Kondo model.   The CFT solution of this problem bears a deep connection to the RR states.  In Sec.~\ref{sec:res_level} we describe the physics of a resonant level tunnel coupled to the edge of a RR state and map it to the $k$-channel Kondo model using the results of Sec.~\ref{sec:CFT_Kondo}.  We also discuss the tunneling of an electron {\em through} the dot. Finally, in Sec.~\ref{sec:discussion} we remark on the experimental implementation of our scenario and summarize in Sec.~\ref{sec:summary}.

\section{Edge theories of Quantum Hall States}
\label{sec:edge_theory}

Fractional quantum Hall states are topologically-ordered\cite{Wen} states of matter: They are gapped in the bulk and contain quasi-particle excitations with non-trivial charge and braiding statistics.  Broken time reversal symmetry in quantum Hall systems leads to net flow of charge in one direction along the boundary.  In the absence of edge reconstruction\cite{Chklovskii:prb92,Chamon:prb94,Wan:prl02,Joglekar:prb03,Orgad:prl08}  the number of independent edge excitations and their low energy properties are determined by the topological order of the bulk state.\cite{Wen}  In this way, there is a ``bulk-edge" correspondence in quantum Hall systems.\cite{Nayak:rmp08,Read07}  CFT can provide a precise description of this correspondence, but note Refs.~[\onlinecite{Bernevig:prl08,Jolad:prl09}].  A given CFT can be used to
produce trial bulk wavefunctions in $2$ spatial dimensions
with and without quasi-particles.\cite{Moore:npb91,Nayak96c,Read:prb99,Bernevig:prl09,Hansson:prl09,Hansson:prb07}
The same CFT can also be used to describe the dynamics of edge excitations in
$1+1$ dimensions. Of course, whether a particular CFT is chosen by nature ultimately depends on whether the corresponding state has a lower energy than other candidate states.  Fortunately, some of the interesting CFTs (namely the $\mathbb{Z}_k$ parafermion theories) do seem to be favored or at least very competitive at certain filling fractions (typically for $2 < \nu <3$) and with reasonable interactions. \cite{Read:prb99,Rezayi:prb09,Wang09}

Our analysis of the coupling between the edge of a non-Abelian fractional quantum Hall state and a quantum dot relies heavily on the theory of the edge.  We focus on the edges of the RR states and their particle-hole conjugates for general $k\geq 2$, which contain the physics of the chiral $\mathbb{Z}_k$ parafermion CFT\cite{CFT} in addition to the physics of the Abelian quantum Hall states.  In this section we present a self-contained summary of the most important aspects of the edge needed for our study.  In order to begin our narrative on familiar ground, we start the discussion by reviewing the essential properties of the edge states of the more widely known Abelian quantum Hall states.  In order to simplify our presentation, we will assume throughout this paper that edge reconstruction does not occur.\cite{Chamon:prb94,Wan:prl02,Joglekar:prb03,Orgad:prl08}  We will discuss the qualitative effects of edge reconstruction in Sec.~\ref{sec:discussion}.

\subsection{Laughlin and Hierachy edge states}
In the integer quantum Hall effect, where electron-electron interactions can be neglected for many purposes, gapless edge excitations occur whenever a Landau level is pushed through the Fermi energy near a boundary.\cite{Halperin:prb82} A quantum Hall state of integer filling $\nu=n$ has $n$ chiral one-dimensional (Dirac) fermionic modes that propagate along the boundary.  These chiral modes carry heat and charge along the edge.  A single chiral Dirac fermion has central charge $c=1$, so for filling fraction $n$, the total effective central charge is $c=n$.  Since the Hall conductivity, $\sigma_H=n \frac{e^2}{h}$, and the thermal conductivity, $\kappa_H=n \frac{\pi^2k_B^2}{3h} T$, in dimensionless units the Hall conductivity and the thermal Hall conductivity are equal.  Here $e$ is the charge of the electron, $h$ Planck's constant, $k_B$ Boltzmann's constant, and $T$ the temperature.

Qualitatively, the situation is similar in an Abelian fractional quantum Hall state: The bulk is gapped and the edge contains an integer number of chiral gapless edge modes.  However, due to the central role that interactions play in the fractional quantum Hall effect, these edge modes carry charges which are a fraction of the electron charge.  The simplest example is the Laughlin state at filling fraction
$\nu=1/m$, with $m$ an odd integer. A Laughlin state contains a single edge mode.  Wen demonstrated\cite{Wen:prl90,Wen:prb90} that the universal properties of this edge mode are described by a chiral Luttinger liquid theory (a Guassian CFT) with edge mode charges and correlations determined by $\nu$ ($\nu=1$ is a special case of this theory).  Since a Gaussian CFT has $c=1$, and the Laughlin state has only one edge mode, the total central charge of the edge is $c=1$.  It should be recalled that while the Hall conductance depends on the value of the edge charge,\cite{Laughlin83} the thermal Hall conductance does not:  all Laughlin states and the $\nu=1$ integer quantum Hall state have identical $\kappa_H$.  Meanwhile, ${\sigma_H}=\nu\,\frac{e^2}{h}$ differs for these states. 

The low-energy edge theory of the Laughlin state (and $\nu=1$) can be expressed in terms of bosonic fields and has an action that takes the form
\begin{equation}
\label{eq:S_L}
S_{\rm edge}^{\rm Laugh}=\frac{1}{4\pi \nu}\int d\tau dx\, \partial_x \phi (i\partial_\tau   + v \partial_x) \phi ,    
\end{equation}
where the chiral bosonic fields satisfy the commutation relations $[\phi(x'),\phi(x)]=i \pi \nu {\rm sgn}(x-x')$; and $v$ is the velocity of the edge mode, determined by non-universal properties of the edge confining potential and interactions. (Throughout our paper we have set $\hbar=1$, which multiplies the velocity $v$ here and in most places in the text.) The electron operator on the edge is 
\begin{equation}
\label{eq:Psi_Laugh}
\Psi_{e,{\rm Laugh}}^\dagger=e^{i\phi/\nu},
\end{equation}
and the quasi-particle operator of charge $\nu e$ is $\Psi_{qp,{\rm Laugh}}^\dagger=e^{i\phi}$.

Wen also argued\cite{Wen92,Wen:ijmp92,Wen95} that the same structure appears in a general hiearchy state\cite{Halperin84,Haldane83} but with multiple chiral edge modes.\cite{MacDonald:prl90,Wen:prl90}  This more general theory of the Abelian quantum Hall hierarchy states takes the form
\begin{multline}
\label{eq:S_Hier}
S_{\rm edge}^{\rm Hier}=\frac{1}{4\pi}\int d\tau dx [i\partial_\tau \phi_i K_{ij} \partial_x \phi_j +  \partial_x \phi_i V_{ij} \partial_x \phi_j\\
+ 2 A_\mu {t_i} \epsilon_{\mu\nu}\partial_\nu {\phi_i}
 ]. 
\end{multline}
The $n\times n$ matrix $K_{ij}$ and the $n$-component vector
$t_i$ encode the topological properties
of the quantum Hall state.
For instance, $K_{ij}=\delta_{ij}+2p$, when the filling fraction is
of the form $\nu =\frac{n}{2pn+1}$ (the Jain sequence \cite{Jain89})
with $n$ and $p$ integers. 
The  $n\times n$ matrix $V_{ij}$ describes the non-universal velocities
and interactions between the $n$ edge modes.
An arbitrary excitation on the edge is created by an
operator $\hat T_{\{m_i\}}(x)=e^{i\sum_{j=1}^n m_j \phi_j(x)}$ where $m_j$ are arbitrary integers.\cite{Wen:ijmp92}
The topological properties of the state are:
(i) Ground state degeneracy on surface of genus $g$ is $\rm{Det}(K)^g$
where $\rm{Det}$ is the determinant.
(ii) Filling fraction $\nu=\sum_{ij} t^\dagger_iK_{ij}^{-1}t_j$.
(iii) Statistical angle of a particle created by $\hat T_{\{m_i\}}$ is
$\frac{\Theta}{\pi}=\sum_{ij}m_i   K_{ij}^{-1}m_j$. (iv) Charge
of a particle created by $\hat T_{\{m_i\}}$
$Q=\sum_{ij}m_i   K_{ij}^{-1}t_j$.
Note that both $K_{ij}$ and $t_i$ are basis-dependent,
and an $SL(n,Z)$ transformation can transform them between
different bases. In the `symmetric basis', in which $t_i=1$ for all $i$,
the electron operator takes the form 
\begin{equation}
\label{eq:Psi_Hier}
\Psi_{e,{\rm Hier}}^\dagger=e^{i\sum_{j=1}^n m_j \phi_j(x)},
\end{equation} 
  where $m_i=\sum_j K_{ij}l_j$, $\sum_i l_i=1$, and $l_j=\sum_i m_i K_{ij}^{-1}$.  This ensures the electron has a statistical angle of $\pi\; (\rm mod\; 2\pi)$ and a unit charge.    It can readily be checked that the formulas above reduce to the correct expressions in a Laughlin state where $K_{ij}\to 1/\nu=m$ and there is only one field $\phi$ in the edge theory.  In the hierarchy states, $\kappa_H=n\frac{\pi^2k_B^2}{3h} T$ when
 all $n$ edge modes move in the same direction.  Otherwise, $\kappa_H=
 (c_{-})\frac{\pi^2k_B^2}{3h} T$, with $c_{-}$ equal to the difference of the central
 charges of the right and left moving edge modes.  If $c_{-}<0$, heat flows in the direction opposite the charge.  This happens, for example, in the $\nu=3/5$ state.

A few remarks are in order regarding the edge theories of Abelian fractional quantum Hall states.  First, we note that {\em all} edge channels in the theory \eqref{eq:S_Hier} carry heat in the theory described above. In the `symmetric basis' mentioned
above, all edge channels carry charge as well. (Depending on the non-universal
matrix $V_{ij}$, the symmetric basis or, perhaps, another basis may be
the eigenmodes of the Hamiltonian.)
The edge theories of non-Abelian fractional quantum Hall states always contain additional electrically neutral modes that carry no charge but do carry heat.  These neutral modes are responsible for the non-Abelian properties of the quantum Hall state.  In the class of quantum Hall states we consider in this paper, the neutral modes will correspond to either Majorana fermions, or their generalization, parafermions (to be described below).  

Second, disorder at the edge plays an important role for fractional quantum Hall states with counter-propagating edge modes.\cite{MacDonald:prl90,Wen:prl90}   If all the charged modes are propagating in the same direction, disorder has no effect at the boundary because edge modes have no states into which they can backscatter.  Forward scattering is unimportant as it has no qualitative effect on the physics.  On the other hand, if the edge contains counter-propagating modes it is possible for backscattering to occur in the presence of disorder.  As is well known, backscattering in one-dimensional interacting systems can lead to dramatic effects.\cite{Kane:prb92} One can then analyze the disorder terms within a renormalization group approach.  A new effective edge theory can emerge at low energies when disorder is relevant.\cite{Kane94,Kane:prb95}  Whether a given state in the hierarchy has counter-propagating modes can be determined simply by looking at the sign of the eigenvalues of the matrix $K$ appearing in Eq.\eqref{eq:S_Hier}.  If there are eigenvalues of opposite sign, there are counter-propagating modes.  This situation occurs at filling fraction $\nu=2/3$ for which $n=2$ and $p=-1$, for example.\cite{MacDonald:prl90,Wen:prl90}    As a rule of thumb,  particle-hole conjugate states contain counter-propagating edge modes and one needs to consider the effect of edge disorder. (The $\nu=2/3$ state can be thought of as $\nu=1/3$ fluid of holes in a $\nu=1$ background. Thus, the edge theory in the absence of disorder can be described by an ``outer" $\nu=1$ edge and an ``inner" $\nu=-1/3$ counter propagating edge.)  The same scenario will arise in particle-hole conjugate states in non-Abelian quantum Hall systems.  One must analyze the relevance of disorder there as well and determine the new effective low-energy theory of the edge if disorder is relevant.\cite{Levin:prl07,Lee:prl07,Bishara:prbR08}

\subsection{Moore-Read Pfaffian edge state}

Fermi statistics requires that a large class of Abelian states
have odd-denominator filling fractions, such as the most prominently observed
ones, which have $\nu=\frac{n}{2pn+1}$. \cite{Jain89}
It thus came as a surprise when a Hall plateau at filling fraction $\nu=5/2$ was observed.\cite{Willett:prl87} Fractional quantum Hall states with
even-denominator filling fractions were proposed, based on the idea
that electrons could form bosonic pairs which could then condense
into a bosonic analogue of the Laughlin state.\cite{Halperin83,Haldane88,Moore:npb91,Greiter92} Such states
could be either Abelian \cite{Halperin83} or non-Abelian.\cite{Moore:npb91}
(It was later realized that one proposed state \cite{Haldane88}
was actually critical.\cite{Read00}) These states are distinguished
by the topological properties of their pair wavefunctions.
The MR Pfaffian \cite{Moore:npb91} state is characterized by
a weakly-bound \cite{Read00} $p_x+ip_y$ pair wavefunction.
(The anti-Pfaffian state is characterized by
a weakly-bound $p_x+ip_y$ pair wavefunction of {\em holes}, or equivalently a $p_x-ip_y$ state of electrons.\cite{Lee:prl07,Levin:prl07})
The MR Pfaffian wavefunction was constructed by adding an electricaly
neutral chiral $\mathbb{Z}_2$ Ising CFT to the $c=1$ chiral boson CFT
that describes the edge of the Laughlin states.
(One typically assumes that the filled lower $\nu=2$ Landau levels
are inert and play no role other than to perhaps modify the effective
electron-electron interactions so that pairing is energetically favorable.)
The MR state, like the Laughlin states, is spin polarized.
Its edge action is the sum of the neutral and charged sectors,
$S=S_n+S_c$ with
\begin{equation}
\label{eq:S_n}
S_n^{\rm MR}=\int d\tau dx \; i\psi \left(i\partial_\tau \psi + v_n \partial_x \psi\right),    
\end{equation}
and
\begin{equation}
\label{eq:S_c}
S_c=\frac{2}{4\pi}\int d\tau dx \partial_x \phi\left(i\partial_\tau \phi  + v_c \partial_x \phi\right) ,    
\end{equation}
where $v_n$ is the neutral mode velocity and $v_c$ is the charge mode velocity. Typically $v_n<v_c$, because of the repulsive interactions in the charge sector.\cite{Wan:prl06,Wan:prb08}   Note that \eqref{eq:S_c} is just \eqref{eq:S_L} with $\nu=1/2$ and $v=v_c$. 

Eq.\eqref{eq:S_n} is equivalent to the right-moving part of the quantum
critical $1+1$-D transverse field Ising model (or the classical critical
$2D$ Ising model). $\psi$ is a Majorana fermion operator. In the Ising
model, it creates a domain wall between up- and down-spins. The Ising model
spin field $\sigma$ is a twist field which creates a branch cut in $\psi$.
The field $\sigma$ is important for understanding non-Abelian statistics
in the MR Pfaffian state and for charge $e/4$ quasiparticle tunneling
from one edge of a droplet to the other. However, we will not be using it in this
paper because we focus here on electron tunneling between one droplet
and another much smaller droplet (i.e. the quantum dot).

The electron creation operator in the MR state is:
\begin{equation}
\label{eq:Psi_MR}
\Psi_{e,{\rm MR}}^\dagger=\psi e^{i2\phi},
\end{equation}
where we have used the convention ${\rm dim}[e^{i\alpha \phi}]=\nu \frac{\alpha^2}{2}$ and ${\rm dim}[\psi]=1/2$, so that ${\rm dim}[\Psi_{e,{\rm MR}}]=3/2$.  Note that the electron operator \eqref{eq:Psi_MR} is a combination of a Majorana fermion and an exponential of the boson, identical to that appearing in \eqref{eq:Psi_Laugh}.  The appearance of a {\em single} Majorana fermion in the electron operator is the crucial factor that allows the mapping of electron tunneling to and from a resonant level to the channel symmetric 2-channel Kondo model.\cite{Fiete:prl08}  Since a Majorana fermion has central charge $1/2$ and, in the MR Pfaffian state,
moves in the same direction as the charge mode,
the total central charge of the edge is $3/2$,
which implies $\kappa_H=\frac{3}{2} \frac{\pi^2k_B^2}{3h} T$.

\subsection{Read-Rezayi edge states}

The Majorana fermion appearing in the theory of the MR state
is just the $k=2$ special case of $\mathbb{Z}_k$ parafermions,
which invites generalizations to higher $k$. These were constructed
by Read and Rezayi;\cite{Read:prb99} they are spin-polarized
states with filling fraction $\nu=k/(k+2)$. They may be applicable
to observed states at $\nu=2+k/(k+2)$ if one, again, assumes
that the filled lower Landau levels are inert. 

The $\mathbb{Z}_k$ parafermion theories arise as the self-dual
critical points of $\mathbb{Z}_k$ clock models.\cite{Fradkin80}
In contrast to the chiral $\mathbb{Z}_2$ Majorana fermions, the $\mathbb{Z}_k$ parafermions\cite{Fateev85} for general $k>2$ are not free theories.\cite{CFT}
They have central charge $c=\frac{2k-2}{k+2}$
and can be realized as SU(2)$_k$/U(1) cosets.
This coset can be realized at the Lagrangian level
by an SU(2)$_k$ chiral WZW model in which
the U(1) subgroup has been gauged:\cite{Karabali90}
\begin{multline}
\label{eq:S-RR}
S_{{SU(2)_k}/U(1)} =  \frac{k}{16\pi}\int d\tau dx\,
\text{tr}\left( {\partial_x} {g^{-1}}
\overline{\partial} g\right)
\\ -\:
i\frac{k}{24\pi}\int dx d\tau dr\,  \epsilon^{\mu\nu\lambda}
\text{tr}\left( {\partial_\mu} g\,{g^{-1}}
{\partial_\nu} g\,{g^{-1}}\,{\partial_\lambda}g\,{g^{-1}}\right)
\\
 + \frac{k}{4\pi} \int dx d\tau \, \text{tr}
\Bigl({A_x}\overline{\partial}g \cdot g^{-1}
- \overline{A}g^{-1}{\partial_x}g
+ {A_x}g  \overline{A}g^{-1}
- {A_x} \overline{A} 
\Bigr).
\end{multline}
In this expression,
$\overline{\partial}\equiv i{\partial_\tau} + {v_c}{\partial_x}$
and $\overline{A}\equiv A_{\tau}-i{v_n}{A_x}$.
The field $g$ takes values in SU(2).
The second integral is over any three-dimensional manifold $M$
which is bounded by the two-dimensional spacetime of the edge $\partial M$.
The value of this integral depends only on the values of the field $g$
at the boundary $\partial M$. The gauge field has no Maxwell term,
so its effect is to set to zero the U(1) current to which it is coupled.

The coset construction can be realized at the operator level
by writing the SU(2)$_k$ currents in terms of a chiral
boson $\phi$ and a parafermion field $\psi_1$:
\begin{eqnarray}
\label{ParafCurrents}
J^+ &=& \sqrt{k}\psi_1 e^{i\phi},\cr
J^{-} &=& \sqrt{k}\psi_1^{\dagger} e^{-i\phi},\cr
J^z &=& \frac{k}{2}\partial_x \phi,
\end{eqnarray}
where the chiral boson is normalized so that
${\rm dim}[e^{i\phi}]=1/k$. The parafermion field is
what is left over after the chiral boson has been
`removed' from the SU(2) current. Throughout this paper different normalization of the bosonic fields will be used as a consequence of the different charge sector Hamiltonians at different filling fractions, but the scaling dimension of the SU(2) currents will always remain unity.  There are also other primary
fields in the $\mathbb{Z}_k$ parafermion theory, such as
powers of the spin field. Although they are important for {\em quasiparticle}
tunneling from one edge of a droplet to the other through the bulk,
they do not enter our analysis here for electrons, where the parafermion field
is central.

The charge sector of the RR edge theory (neglecting the 2 filled lower Landau levels) is identical to \eqref{eq:S_L} with $\nu=k/(k+2)$. 
The electron operator in the RR state is obtained by combining
appropriate fields from the charge and neutral sectors:\cite{Read:prb99}
\begin{equation}
\label{eq:Psi_RR}
\Psi_{e,{\rm RR}}^\dagger=\psi_1 e^{i\frac{k+2}{k}\phi},
\end{equation}
which has ${\rm dim}[\Psi_{e,{\rm RR}}]=3/2$, independent of $k$.  With the
$c=1$ charge sector included, the total central charge of the edge is
$\frac{3k}{k+2}$, which implies $\kappa_H=\frac{3k}{k+2} \frac{\pi^2k_B^2}{3h} T$ in the RR states.

\subsection{Particle-hole conjugates of MR and RR states} 

Particle-hole conjugate fractional quantum Hall states can be thought of as a Hall fluid of holes in a background of electrons at integer filling. \cite{MacDonald:prl90,Wen:prl90}  This
picture naturally leads to an ``outer edge" with properties like that of an integer quantum Hall state, and an ``inner edge" (that propagates in the opposite direction because of the opposite charge of the holes relative to electrons) that is at a filling $1-\nu$,
where $0<\nu<1$ is the filling fraction of the state that is conjugated.  (We have assumed the particle-hole conjugation is in the lowest Landau level by again neglecting the lower 2 filled Landau levels.  If the filling fraction is in the range $2 < \nu < 3$, then the particle-hole conjugate state would be at filling fraction $5-\nu$.)  As noted in the
introduction, the MR Pfaffian state $\nu=5/2$ has a particle-hole conjugate state (degenerate in the absence of Landau level mixing, which is a particle-hole
symmetry breaking perturbation) that is at the same filling fraction.\cite{Levin:prl07,Lee:prl07}
This immediately raises an additional question: ``Is the MR Pfaffian state or its particle-hole conjugate a better candidate for $\nu=5/2$?" Our work provides an experimental test for this question quite distinct from existing proposals.

As we mentioned in our discussion of the hierarchy states, the edge physics is more subtle for states that have counter-propagating modes on the edge.  Much of the non-trivial physics of one-dimension is related to backscattering in the presence of interactions.  Counter-propagating modes allow for such backscattering, provided the disorder causing the backscattering has Fourier components at the right values to soak up the momentum mismatch between the two (or more) modes.  The relevant analysis in the fractional quantum Hall context was first carried out by Kane {\it et al.} for the $\nu=2/3$ state.\cite{Kane94}  The key result is that the edge has two phases: one in which the disorder
is a relevant perturbation to the ``clean" edge theory and one in which it is not.
When the disorder is irrelevant, the picture of an outer edge with properties of an integer quantum Hall state and an inner counter-propagating edge with properties of a fractional quantum Hall state applies.  On the other hand, when disorder is relevant [which,
in turn, depends on the non-universal matrix $V_{ij}$ in \eqref{eq:S_Hier}], the edge reconstructs at long length scales into an electrically neutral (disorder dependent) mode and a Laughlin-like charge mode effectively at $\nu=2/3$.  A remarkable result is that the neutral mode has an emergent SU(2) symmetry.\cite{Kane94}  This neutral mode propagates in the opposite direction to the charge mode.

Much of the analysis of Kane {\it et al.} for the $\nu=2/3$ state carries over to the particle-hole conjugates of the MR and RR states, although the technical details differ somewhat: the analysis for the non-Abelian states needs to be modified to take into account the neutral $\mathbb{Z}_k$ chiral parafermion mode present in the inner edge.\cite{Levin:prl07,Lee:prl07,Bishara:prbR08}  When this analysis is carried out, the edge theory of the particle-hole conjugate of the MR state, ${\overline {\rm MR}}$, is given by $S=S_n^{\overline{\rm MR}}+S_c$ with $S_c$ given by \eqref{eq:S_c} and a neutral sector of 3 flavors of Majorana fermions
\begin{eqnarray}
\label{eqn:three-Majoranas}
S_n^{\overline{\rm MR}}
= \sum_{a=1}^3 \int dx \,i{\psi_a}(-i{\partial_\tau}+
v_n{\partial_x}){\psi_a},
\end{eqnarray}
that propagates in the direction opposite of the charge.  Since the central charge of a Majorana fermion is 1/2 and there are three flavors of them, $\kappa_H=-\frac{1}{2} \frac{\pi^2k_B^2}{3h} T$, so that heat flows ``upstream" relative to the current.  (Of course, once the 2 lower Landau levels are taken into account, heat will still flow downstream along with the current, but there will be a reduction relative to the MR state.)  The different edge theories  and thermal Hall conductivities for the MR and ${\overline {\rm MR}}$ states imply that these are topologically distinct states.\cite{Levin:prl07,Lee:prl07}
The neutral sector \eqref{eqn:three-Majoranas} is SU(2) symmetric and, in fact,
the symmetry generators,
\begin{equation}
J^a = i \epsilon^{abc} \psi_b \psi_c,
\end{equation}
obey an SU(2)$_2$ Kac-Moody algebra, i.e.
\begin{equation}
{J^a}(z) \cdot {J^b}(0) = \frac{1}{z^2} \,\frac{k}{2} \delta^{ab} + \frac{1}{z}\,i\epsilon^{abc}
{J^c}(0) + \ldots
\end{equation}
with $k=2$.
By contrast, in the $\nu=2/3$ case analyzed by
Kane {\it et al.} \cite{Kane94}, the neutral sector can be formulated as an SU(2)$_1$
Kac-Moody algebra.

The electron operator in the ${\overline {\rm MR}}$ state is no longer unique as it was in the MR state and this will ultimately lead to different Kondo physics when the resonant level problem is mapped to the Kondo model.  There are three different dimension-$3/2$ (the smallest possible scaling dimension) electron operators,\cite{Levin:prl07,Lee:prl07}
\begin{equation}
\left(\Psi_{e,{\overline {\rm MR}}}^\dagger\right)_a={\psi_a} e^{2i\phi},
\end{equation}
for $a=1,2,3$ where the combination $({\psi_1} - i {\psi_2})e^{2i\phi}$ is inherited from the
electron operator of the $\nu=1$ integer quantum Hall state in which a Pfaffian state of holes forms.
The other two electron operators are complicated charge-$e$ combinations of the $\nu=1$ electron operator and particle-hole excitations between the inner MR and outer
$\nu=1$ edges. The  3 flavors of Majorana fermions present in the ${\overline {\rm MR}}$ state will turn out to drive the stable 2-channel Kondo physics of the MR state to the less exotic single-channel Kondo physics in the ${\overline {\rm MR}}$ state.\cite{Fiete:prl08}

The edge theory of the ${\overline {\rm RR}}$ state for general $k\geq 2$ is a straightforward generalization of the ${\overline {\rm MR}}$ state:  the neutral sector is described by an SU(2)$_k$ theory and the charge sector is given by \eqref{eq:S_L} with $\nu=2/(k+2)$.  The counter-propagating neutral sector is therefore given by a chiral WZW theory for SU(2) at level $k$,
\begin{eqnarray}
\label{eq:S_n_k}
S_{{\rm WZW},k}&=&\frac{k}{16}\int d\tau dx {\rm tr}(\partial_x g^{-1}\overline\partial g)\nonumber\\
&-i& \frac{k}{24\pi}\int dr d\tau dx \epsilon^{\mu \mu \lambda} {\rm tr}( \partial_\mu g g^{-1} \partial_\nu g g^{-1} \partial_\lambda g g^{-1}),\nonumber \\
\end{eqnarray}
The central charge of the neutral mode is $\frac{3k}{k+2}$, so the net thermal
conductivity is $\kappa_H=-2\left(\frac{k-1}{k+2}\right)\frac{\pi^2k_B^2}{3h} T$.
Since $k\geq 2$, heat flows in the opposite direction of charge.

As in the case of the ${\overline {\rm MR}}$ state, the electron operator is no longer unique.\cite{Bishara:prb08,Bishara:prbR08}  There are $2k+1$ electron operators with left-and right-scaling
dimensions $\Delta_R = (k+2)/4$, $\Delta_L=k/4$, so that their total scaling dimension
is $(k+1)/2$  and their conformal spin is $\Delta_R - \Delta_L= 1/2$,
as required for a fermion. These operators are given by\cite{Bishara:prbR08}
\begin{equation}
\label{eq:el_RRbar}
\left(\Psi_{e,{\overline {\rm RR}}}^\dagger\right)_m=\chi_{j=k/2}^{m} e^{i\frac{k+2}{2}\phi},
\end{equation}
with $m=-k/2, -k/2+1,...,k/2$.  The $\chi_{j=k/2}^{m}$ are related to the spin $j=k/2$  primary fields of SU(2)$_k$ and are built entirely from fields in the neutral sector of the edge theory.  The $\chi_{j=k/2}^{m}$ have scaling dimension $k/4$ and can be constructed by operating with the SU(2)$_k$ current operator $J^+\sim \psi_1 e^{i\phi_\sigma}$ multiple times on a ``bare outer edge" electron operator.\cite{Bishara:prbR08}  Here $\phi_\sigma$ is an electrically neutral combination of the ``inner" and ``outer" charge edge modes\cite{Bishara:prbR08} and $\psi_1$ is the  $\mathbb{Z}_k$ parafermion.  The non-uniqueness of the electron operator \eqref{eq:el_RRbar} will again turn out to drive the system away from the channel symmetric $k$-channel fixed point.

To summarize, the neutral sectors of the MR Pfaffian
and RR edges are the Ising model (Majorana fermion)
and $\mathbb{Z}_k$, $k\geq 3$ parafermion theories, respectively.
These neutral sectors can be realized as the coset
SU(2)$_k$/U(1) ($k=2$ is the Ising case). In contrast, the
neutral sectors of the anti-Pfaffian and $\overline{\rm RR}$
edges are oppositely-directed SU(2)$_k$ theories.
In effect, particle-hole conjugation undoes the coset and
restores the full SU(2)$_k$ Kac-Moody algebra (in the phase
in which disorder is relevant, so that the edges
equilibrate).

\section{The $k$-channel Kondo Model}
\label{sec:CFT_Kondo}

In this section we describe the $k$-channel Kondo model and its solution using analytical methods that we will encounter in the next section where our scenario of a quantum dot near a bulk non-Abelian quantum Hall state (see Fig.\ref{fig:schematic}) is discussed.  The most familiar guise of the Kondo model is as a local magnetic impurity (spin) in a sea of conduction electrons.  Electrons scattering from this local moment can flip their spin. As a result, at low energies non-trivial correlations develop between the magnetic moment and the electrons of the many-body fermi sea, and the local moment is screened (or perhaps ``over" or ``under" screened).\cite{Nozieres:jpp80,Fiete:prb02}  Describing this behavior in detail has been one of the most intensively studied problems in condensed matter physics over the last 40 years.\cite{Kondo}  Research on the problem is still quite active, particularly as it  relates to finite frequency or non-equilibrium properties.  The mapping we establish in this work may provide another venue to probe interesting non-equilibrium properties of a Kondo system, particularly in the multi-channel case, which would be
an interesting route for future study.

\subsection{The $k$-channel Kondo Hamiltonian}

Our results only require equilibrium properties which are well understood for the cases of interest to us.  The most important distinction that needs to be made is between single channel and multi-channel versions of the Kondo model.\cite{Nozieres:jpp80} Here the ``channels" refers to the number of ``flavors" of electrons that couple to the impurity spin. 

In the Kondo problem, one considers $k$ channels of
conduction electrons (e.g. $k$ bands) interacting with
a localized impurity spin $\vec{S}$, which, without loss of
generality, we can take to be at the origin. The impurity
spin is treated as point-like, so that only the $s$-wave component
of the conduction electrons interacts with it. By restricting to
the $s$-wave channel, we effectively make the problem
one-dimensional. The radial dimension can be treated
as running from $r=0$ to $r=\infty$, with $k$ species of
non-chiral Dirac fermions $\psi_{i}^{R,L}(r)$, $i=1,2,\ldots,k$
on the half-line. They do not interact with each
other but interact with a spin $\vec{S}$ at $r=0$. Alternatively,
we can formulate the model in terms of $k$ species of
chiral Dirac fermions on the line full line by allowing $r$
(which we rename as $x$) to run from $-\infty$ to $\infty$ 
and flipping the right-moving fermions onto the negative axis
$\psi_{i}^{L}(-x)\equiv\psi_{i}^{L}(x)$ for $x>0$.
Thus, the $k$-channel Kondo Hamiltonian has
$k$ chiral Dirac fermions, interacting with the spin
\begin{eqnarray}
\label{eq:H_K}
H_K=\sum_{i=1}^k \Biggl\{H_0[\psi_i]+\frac{\lambda_{\perp i}}{2}(J_i^+(0)S^-+{\rm H.c})\nonumber \\+\lambda_{z i}J_i^z(0)S^z\Biggr\} +hS^z,
\end{eqnarray}
where $H_0$ describes the $k$-channels of the conduction
electrons $\psi_i$ (not parafermions!)
\begin{equation}
H_0[\psi] = \int dx\, {v_F}\, \psi^\dagger i{\partial_x} \psi.
\end{equation}
The local moment $\vec S$ is situated at the origin.  The local conduction electron spin density in channel $i$ is given by 
\begin{eqnarray}
J_i^-(x) &=& :\psi_{i\downarrow}^\dagger(x)\psi_{i\uparrow}(x):,\cr
J_i^+(x) &=& :\psi_{i\uparrow}^\dagger(x)\psi_{i\downarrow}(x):,\cr
J_i^z (x)&=& \frac{1}{2} : [\psi_{i\uparrow}^\dagger(x)\psi_{i\uparrow}(x)-\psi_{i\downarrow}^\dagger(x)\psi_{i\downarrow}(x)]:,
\end{eqnarray}
where $:\;\;:$ denotes normal ordering.  The model is said to be {\em exchange} isotropic if $\lambda_{\perp i}=\lambda_{z i}$, and {\em channel} symmetric if $\lambda_{\perp i}=\lambda_{\perp}$ and $\lambda_{z i}=\lambda_{z}$ for all channels $i$.  The last term in \eqref{eq:H_K} is present if there is an externally applied magnetic field along the $z$-direction.

The ground state properties of \eqref{eq:H_K} generally depend on $k$, the size of the spin $\vec S$; the value of the external field $h$; the presence/absence of anisotropy in the exchange and asymmetry
in the channel coupling; and the signs of the products $(\lambda_{\perp i})^2 \lambda_{zi}$.  Fortunately, we need only consider a small subset of the possible cases.   From the point-of-view of ``exotic" physics, one of the most important cases is when $S=1/2$, $h=0$, and the model is {\em channel symmetric} (exchange anisotropy is unimportant, {\it i.e.} irrelevant\cite{Affleck:prb92} in the renormalization group sense).  In this case, the ground state exhibits non-Fermi liquid properties for $k\geq 2$ in the channel symmetric case.
However, these non-Fermi liquid fixed points are unstable to
channel asymmetry, which drives the system away from the channel symmetric limit (and to the $k=1$ Fermi liquid fixed point). We note that in our scenario discussed in Sec. \ref{sec:res_level} and shown in Fig.~\ref{fig:schematic}, the ``$k$" appearing here is exactly the same ``$k$" that appears in the RR states at filling fraction $\nu =\frac{k}{k+2}$ or the $\overline {\rm RR}$ states at filling fraction $\nu =\frac{2}{k+2}$. Furthermore, as we will see, our model automatically maps to the channel symmetric case.

\subsection{Emery-Kivelson solution of $H_K$  for $k=2$ and $S=1/2$}

The Emery-Kivelson (EK) analysis\cite{Emery:prb92} of the 2-channel $S=1/2$ Kondo model plays a central role in our work, so we briefly review it here.  EK solved the 2-channel Kondo model by mapping it onto a Majorana resonant level model. 
(This gives a clue to the relation with our problem
since Majorana fermions are an important part of the edge
theories of the MR Pfaffian state and the anti-Pfaffian state.)
They studied the properties of the resonant level model and then mapped them back onto the original Kondo model to determine quantities such as the impurity susceptibility and ground state entropy.  In Sec.~\ref{sec:res_level} we will do the reverse: start with a particular resonant level model and ask ``To what impurity (Kondo) model does it correspond?"  We note, however, that not all resonant level models immediately map to a Kondo model, as we show explicitly in the case of a quantum dot coupled to {\em two} bulk fractional quantum Hall reservoirs. 

The EK solution begins by bosonizing the electron operators \cite{Emery:prb92} 
\begin{equation}
\psi^\dagger_{i\sigma}(x)=\frac{1}{\sqrt{2\pi a_0}}\, e^{i\phi_{i\sigma}(x)},
\end{equation}
where $\sigma=\uparrow,\downarrow$,
In terms of these bosons, the local conduction electron spin density is:
\begin{eqnarray}
J_i^+(x)=\frac{1}{2\pi a_0}\,e^{i\sqrt 2\phi_{i-}(x)}, \;\;
J_i^z(x)=\frac{1}{\sqrt 2\pi} \partial_x \phi_{i-}(x),\,
\end{eqnarray}
where
\begin{equation}
\phi_{i\pm}(x)=\frac{1}{\sqrt 2} [\phi_{i\uparrow}(x)\pm
\phi_{i\downarrow}(x)],
\end{equation}
$a_0$ is a short distance cut-off.  We keep our normalization
convention from before so that the electron operator has scaling dimension 1/2 and the currents have scaling dimension 1. 
Focusing on the 2-channel case ($i=1,2$), EK then introduce
``spin" and ``spin-flavor" bosons as the following
linear combinations
\begin{eqnarray}
\phi_{s}(x)=\frac{1}{\sqrt 2} [\phi_{1-}(x)+\phi_{2-}(x)],\\
\phi_{sf}(x)=\frac{1}{\sqrt 2} [\phi_{1-}(x)-\phi_{2-}(x)],
\end{eqnarray}
which, as we will see momentarily, couple to the impurity spin
while the fields
\begin{eqnarray}
\phi_{c}(x)=\frac{1}{\sqrt 2} [\phi_{1+}(x)+\phi_{2+}(x)],\\
\phi_{f}(x)=\frac{1}{\sqrt 2} [\phi_{1+}(x)-\phi_{2+}(x)],
\end{eqnarray}
are not coupled to it.
In terms of these fields, the Hamiltonian \eqref{eq:H_K}
takes the form,
\begin{multline}
H_K = H_0[\phi_c]+H_0[\phi_{f}]+H_0[\phi_s]+H_0[\phi_{sf}]\,+\\
\Bigl (\frac{e^{i\phi_s(0)}}{2\pi a_0}\Bigl\{\lambda_{\perp +} \cos[\phi_{sf}(0)]
+i\lambda_{\perp -}\sin[\phi_{sf}(0)]\Bigr \}S^- +{\rm H.c.}\Bigr)\\
+\frac{1}{\sqrt \pi}[\lambda_{z+}\partial_x\phi_s(0)+\lambda_{z-}\partial_x \phi_{sf}(0)]S^z+hS^z,
\end{multline}
where
\begin{equation}
\lambda_{\perp \pm}=\frac{1}{2}\left(\lambda_{\perp 1}\pm  \lambda_{\perp 2}\right),\;\;\;
\lambda_{z \pm}=\frac{1}{2}\left(\lambda_{z 1}\pm  \lambda_{z 2}\right).
\end{equation}
Since $\phi_c$ and $\phi_f$ do not couple to the impurity spin,
we ignore them from now on.

In order to map the Kondo Hamiltonian onto a resonant level model coupled to free fermions, EK perform a unitary transformation
generated by
\begin{equation}
\label{eqn:EK-Unitary}
U=e^{i\alpha S^z \phi_s(0)},
\end{equation}
in order to decouple $\phi_s$
and thereby change the scaling dimension of the transverse exchange
coupling to 1/2.  Under the unitary transformation
generated by (\ref{eqn:EK-Unitary}),
\begin{eqnarray}
H_K\to U^\dagger H_K U=H_0[\phi_s]+H_0[\phi_{sf}] \hspace{2cm}\nonumber \\
+ \Bigl (\frac{e^{i(1-\alpha)\phi_s(0)}}{2\pi a_0}\Bigl\{\lambda_{\perp +} \cos[\phi_{sf}(0)] \hspace{1cm}\nonumber \\
+i\lambda_{\perp -}\sin[\phi_{sf}(0)]\Bigr \}S^- +{\rm H.c.}\Bigr)\nonumber \\
+\frac{1}{\sqrt \pi}[\tilde \lambda_{z+}\partial_x\phi_s(0)+\lambda_{z-}\partial_x \phi_{sf}(0)]S^z +hS^z,
\end{eqnarray}
where
\begin{equation}
\tilde \lambda_{z +}=\lambda_{z +}- \alpha v_F.
\end{equation}
By choosing $\alpha=1$, one eliminates the coupling of the $\phi_s$ field to $S^\pm$ and sets the scaling dimension of the transverse exchange coupling to 1/2 so that the $\phi_{sf}$ sector can be refermionized,
\begin{equation}
\psi_{sf}^\dagger(x)=e^{i\pi d^\dagger d}\frac{1}{\sqrt{2\pi a_0}}e^{i\phi_{sf}(x)},
\end{equation}
where a Klein factor $e^{i\pi d^\dagger d}$ is inserted to ensure the proper anticommutation relations between the $\psi_{sf}(x)$ fermions and the fermions introduced to describe the local magnetic moment,
\begin{equation}
S^+=d^\dagger,\;\;S^-=d, \;\; S^z=d^\dagger d-1/2.
\end{equation}
The operators $d,d^\dagger$
satisfy the anticommutation rules $\{d,d^\dagger\}=1$.  In this representation, the Hamiltonian takes the form
\begin{eqnarray}
\label{eq:H_K_final_gen}
H_K=H_0[\phi_s]+H_0[\psi_{sf}]+h(d^\dagger d-1/2) \hspace{1.5cm} \nonumber  \\
+\frac{\lambda_{\perp +}}{\sqrt{8 \pi a_0}}(\psi_{sf}^\dagger(0)+\psi_{sf}(0))(d-d^\dagger)\nonumber \\
+\frac{\lambda_{\perp -}}{\sqrt{8 \pi a_0}}(\psi_{sf}^\dagger(0)-\psi_{sf}(0))(d+d^\dagger)\nonumber \\
+[\tilde \lambda_{z+}\frac{\partial_x\phi_s(0)}{\sqrt \pi}+\lambda_{z-}: \psi_{sf}^\dagger(0)\psi_{sf}(0):](d^\dagger d-1/2).
\end{eqnarray}
At its generalized Toulouse point,
$\tilde \lambda_{z+}=\lambda_{z-}=0$,
the Hamiltonian \eqref{eq:H_K_final_gen} is quadratic and,
therefore, exactly solvable.\cite{Emery:prb92}

Focusing for the moment on the Toulouse limit, the EK solution proceeds by introducing a Majorana representation for the local moment fermions and the spin-flavor fermions,
\begin{eqnarray}
d&=&\frac{1}{\sqrt 2} (a+ib),\\
\psi_{sf}(x)&=&\frac{1}{\sqrt 2}(\zeta_1(x)+i\zeta_2(x)),
\end{eqnarray}
where $a,b,\zeta_1,\zeta_2$ are Majorana fermions. In this representation, the Hamiltonian (apart from the  degrees of freedom that decouple from the impurity spin) takes the form,
\begin{eqnarray}
\label{eq:H_K_final}
H_K&=&H_0[\zeta_1]+H_0[\zeta_2]+h(iab-1/2)\hspace{1cm}\nonumber\\
&+&i\left(\frac{\lambda_{\perp +}}{\sqrt{2 \pi a_0}}\right)\zeta_1(0)b+i\left(\frac{\lambda_{\perp -}}{\sqrt{2 \pi a_0}}\right)\zeta_2(0)a,
\end{eqnarray}
which describes two Majorana resonant levels ($a$ and $b$) coupled to {\em two} different Majorana modes ($\zeta_1$ and $\zeta_2$) originating from the Fermi sea.  Note that the magnetic field couples the two Majorana resonant levels even at the Toulouse point. 

Let us now focus on \eqref{eq:H_K_final} with $h=0$.  In this case, the Majorana modes completely decouple from each other, and the Kondo Hamiltonian reduces to a sum of 2 Hamiltonians each describing the coupling of a different Majorana resonant level to a different Majorana bath.  It can be shown that the 1-channel Kondo model at its Toulouse point can be mapped into just such a form.\cite{Emery:prb92,Fabrizio:prl95}  (The transverse coupling in the 1-channel model takes the form $\psi^\dagger d+ d^\dagger\psi$ which is equivalent to \eqref{eq:H_K_final} with a constant shift of $\phi_{sf}$.) Therefore, if $\lambda_{\perp +}\neq 0$ and $\lambda_{\perp -}\neq 0$, the 2-channel Kondo model exhibits 1-channel Kondo behavior.\cite{Fabrizio:prl95,Andrei:prl95}
On the other hand, if the model is {\em channel} symmetric ($\lambda_{\perp -}= 0$ and $\lambda_{z-}=0$) then only the ``$b$" Majorana fermion couples to the conduction electrons and non-Fermi liquid properties result.  For example, there is a non-trivial ground state entropy of 
\begin{equation}
S_{\rm imp}(0)=\frac{1}{2}\ln(2),
\end{equation}
 and an impurity susceptibility, 
 \begin{equation}
 \chi_{\rm imp}(T) \propto \ln(T_K/T),
 \end{equation}
  where $T_K$ is the Kondo temperature.\cite{Emery:prb92,Sengupta:prb94}  

Detailed analysis\cite{Emery:prb92,Fabrizio:prl95,Andrei:prl95,Sengupta:prb94,Affleck:prb92} of the full Hamiltonian \eqref{eq:H_K_final_gen}  leads to the following key results: The 2-channel non-Fermi liquid fixed point at $\tilde \lambda_{z +}=\lambda_{z -}=\lambda_{\perp -}=h=0$ is (i) stable to {\em exchange} anisotropy, $\lambda_{\perp+}\neq \lambda_{z+}\neq0$. (ii) unstable to {\em channel} anistropy, $\lambda_{z -}\neq 0$ or $\lambda_{\perp -}\neq 0$. (iii) unstable to a finite magnetic field $h \neq 0$.  In case (ii) the low-energy properties are controlled by the 1-channel Kondo model fixed point (provided the effective coupling is antiferromagnetic) at which the ground state entropy is zero and the impurity susceptibility is a constant.  However, provided the asymmetry is not too large, the 2-channel behavior may survive over a fairly large energy scale before crossing over to the 1-channel behavior at the lowest energies.\cite{Andrei:prl95,Fabrizio:prl95}  For case (iii) the low energy behavior is described by a potential scattering problem with different phase shifts for up- and down-spin electrons.\cite{Affleck:prb92}

\subsection{Affleck-Ludwig solution of $H_K$ for
general $k$ and $S=1/2$}

For $k>2$ channel Kondo models,
the Abelian bosonization method of EK is not as helpful
(except for the $k=4$ channel case\cite{Fabrizio:prl95}),
so other approaches are needed.  For our later discussion, the boundary CFT solution to the $k$-channel (including single-channel) Kondo model  pioneered by Affleck and Ludwig\cite{Affleck:npb90,Affleck:npb91,Affleck_2:npb91} turns out to be particularly useful.  Since the RR and $\overline{\rm RR}$ boundary excitations are described by a CFT (the same CFT, in fact, that appears in the multi-channel Kondo model), this approach provides a natural link between multi-channel Kondo models and non-Abelian quantum Hall states.  The shared CFT is the ``deep" reason we can establish a mapping between the resonant level model at the edge of a RR or $\overline{\rm RR}$ state and the multi-channel Kondo model.  We emphasize that the boundary CFT method covers all cases, even the $(k=1)$-channel Kondo model.

The boundary CFT solution to \eqref{eq:H_K} begins with a conformal decomposition of the conduction electrons.\cite{Affleck:npb90,Affleck:npb91,Affleck_2:npb91}  The conduction electron term represented by $H_0$ is expressed as a sum of three terms (referred to as a ``conformal embedding'') that describe ``charge", ``spin", and ``flavor" sectors.  These sectors posses current operators with U(1), SU(2), and SU($k$) symmetry, respectively.  Specifically, one has
\begin{equation}
\label{eq:H_conf_decomp}
\sum_{i=1}^kH_0[\psi_i]=H[U(1)]+H[SU(2)_k]+H[SU(k)_2].
\end{equation}
The identity \eqref{eq:H_conf_decomp} is typically motivated\cite{Gogolin} by noting that the central charges on the left and right hand sides are equal: $2k=1+\frac{3k}{k+2}+\frac{2(k^2-1)}{k+2}$.
Conformal embedding is useful for the $k$-channel Kondo model because it is only the SU(2)$_k$ currents that couple to the local moment $\vec S$: the conduction electron terms with U(1) and SU($k$) symmetry thus play no role in the impurity physics.  (Recall that the $\mathbb{Z}_k$ parafermions in the RR and $\overline{ \rm RR}$ states are constructed from the coset SU(2)$_k$/U(1). Thus, SU(2)$_k$ plays a key role in both the non-Abelian fractional quantum Hall states of interest here and the $k$-channel Kondo model.)

In terms of the $k$ channels of the conduction electrons, the SU(2)$_k$ spin currents can be expressed as
\begin{equation}
\vec J(x)=\sum_{i=1}^k:\psi^\dagger_{\alpha, i}(x)\frac{\vec \sigma_{\alpha \beta}}{2}\psi_{\beta, i}(x):,
\end{equation}
where $\vec \sigma$ is the vector of Pauli spin matrices.
The Fourier components of the SU(2)$_k$ spin currents satisfy the Kac-Moody algebra,
\begin{equation}
[J_n^a,J_m^b]=i\epsilon^{abc}J^c_{n+m}+\frac{nk}{2}\delta^{ab}\delta_{n+m,0}.
\end{equation}
Thus, the currents can also be expressed in terms of
a parafermion and a boson, as in (\ref{ParafCurrents}).

The next step is to study the coupling of the magnetic impurity to the spin currents.  Let us first assume the coupling of the spin currents to the local moment $\vec S$ is channel symmetric.  The Kondo Hamiltonian is then (neglecting parts that decouple from the spin)
\begin{eqnarray}
H_K=\frac{2\pi v_F}{k+2}\int_0^L dx :\vec J(x)\cdot \vec J(x): \hspace{2.5cm}\nonumber \\
+\lambda_\perp ( {J^+}(0)S^{-} + {J^-}(0)S^{+} )+ \lambda_z {J^z}(0)S^{z} + h {S^z}.
\end{eqnarray}
To illustrate the key features of the Affleck-Ludwig boundary CFT solution it is useful to specialize to the exchange isotropic limit in zero magnetic field.\cite{Affleck:npb90,Affleck:npb91,Affleck_2:npb91}  The
relevance/irrelevance of exchange anisotropy
and channel asymmetry, as well as an externally applied magnetic field, can be analyzed about this point.\cite{Affleck:prb92}  Going to the Sugawara form of the Hamiltonian in Fourier space, we have
\begin{equation}
\label{eq:H_K_Sugawara}
H_K=\frac{2\pi v_F}{L}\left(\sum_{n=-\infty}^\infty \frac{1}{k+2}:\vec J_{-n}\vec J_n:+\lambda_K\sum_{n=-\infty}^\infty \vec J_n \cdot \vec S\right),
\end{equation}
where $\lambda_K$ is the effective Kondo coupling.  The crucial point is that when $\lambda_K=2/(k+2)$ (which corresponds to the low-temperature fixed point), the Hamiltonian \eqref{eq:H_K_Sugawara} becomes equivalent, after completing the square, to an uncoupled spin but with shifted current operators
\begin{equation}
\vec {\cal J}_n=\vec J_n+\vec S.
\end{equation} 
In other words, at the low-energy fixed point, the local moment is ``absorbed" into the spin currents.  This is referred to as the ``fusion hypothesis". In boundary CFT terms, it
can be stated in the following way. Let us fold the system
again, so that we have non-chiral fields on the half-plane
$x>0$, $-\infty<\tau<\infty$.
Each possible conformally-invariant
boundary condition at the boundary of the half-plane,
$x=0$, corresponds to a primary field of the theory
(this is true in a large class of theories, namely those with diagonal partition functions, which includes the models discussed here).
When the interaction is tuned to zero, ${\lambda_K}=0$,
the boundary condition corresponds to the identity operator
in SU(2)$_k$.
According to the ``fusion hypothesis", the boundary condition
at the infrared fixed point is obtained by fusing the
identity with the spin-$1/2$ primary field (or, more generally,
the spin-$S$ primary field), i.e. the boundary condition
is associated with the spin-$1/2$ primary field. This leads
to an entropy {\it gain} of $\ln[2\cos(\pi/(k+2))]$ associated
with the conformal boundary condition. However, since the impurity spin, with its $\ln 2$ entropy is screened, the net entropy change
is $\Delta S = -\ln 2 + \ln[2\cos(\pi/(k+2))]$. If we separate
bulk and impurity so that the initial impurity entropy is $\ln 2$
(how we choose to divide the total entropy of the system
into bulk and impurity is, in part, a convention),
then the entropy in the infrared is:
\begin{equation}
 S_{\rm imp}(0) = \ln[2\cos(\pi/(k+2))].
\end{equation}
We have already seen this scenario in different language
in Emery and Kivelson's solution of the $k=2$ case.
Let us consider the channel symmetric case as the generalized
Toulouse point, so that $\lambda_{\perp-}=0$.
For $\lambda_{\perp+}=0$, the Majorana fermion
$\zeta_1$ in Eq.\eqref{eq:H_K_final} has boundary condition
$\zeta_{1R}(0)=\zeta_{1L}(0)$, which corresponds
to fixed boundary condition in the Ising model. However,
for $\lambda_{\perp+}>0$, which flows to the
infrared fixed point $\lambda_{\perp+}=\infty$,
the boundary condition is $\zeta_{1R}(0)=-\zeta_{1L}(0)$
[as may be seen by direct solution of the quadratic Hamiltonian
\eqref{eq:H_K_final}]. This corresponds to free boundary condition
in the Ising model, which has a boundary entropy which
is higher by $\ln\sqrt{2}$. However, since the impurity
spin's entropy $\ln 2$ is lost, there is a net entropy loss
of $\ln\sqrt{2}$.

As a result of the change of boundary condition in
the SU(2)$_k$ sector while the U(1) and SU(k)$_2$
sectors are unaffected, correlation functions of electron
operators (which combine these three sectors) become non-trivial.
For instance, the impurity susceptibility for the
$k$-channel Kondo model is:
 \begin{equation}
 \chi_{\rm imp}(T) \propto T^{-(k-2)/(k+2)}.
 \end{equation}
The fusion hypothesis has been tested against exact Bethe-ansatz results\cite{Andrei:prl84} and numerical renormalization group (NRG) studies and is now believed to be on quite solid ground.\cite{Affleck:prb92}

The conclusions reached with the boundary CFT analysis\cite{Affleck:prb92} for $S=1/2$ and general $k\geq 2$ are the same
as those reached by the EK analysis for $k=2$
as far as symmetry-breaking perturbations are concerned: The $k$-channel non-Fermi liquid fixed point is: (i) stable to {\em exchange} anisotropy. (ii) unstable to {\em channel} asymmetry. (iii) unstable to a finite magnetic field.  In case (ii) the low-energy properties are controlled by the 1-channel Kondo model fixed point (provided the effective coupling is antiferromagnetic).  For case (iii) the low energy behavior is described by a potential scattering problem with different phase shifts for up and down electrons.\cite{Affleck:prb92}

\section{Exotic Resonant Level Models}
\label{sec:res_level}

Having laid the necessary groundwork for our study in Secs.~\ref{sec:edge_theory} and \ref{sec:CFT_Kondo}, we are now ready to investigate the physics of a quantum dot (modeled by a single
resonant level) coupled by electron tunneling
to a Read-Rezayi quantum Hall state or its particle-hole conjugate.  As we mentioned in the introduction, our main result is that the physics is governed by some version of the multi-channel Kondo model.  

Specifically, we study the set-up shown in Fig.~\ref{fig:schematic}.
The bulk quantum Hall state on the left is assumed to be in a non-Abelian
fractional quantum Hall state at $\nu=2+k/(k+2)$ or $\nu=2+2/(k+2)$.
A quantum point contact may be used to pinch off a finite region of the quantum Hall fluid and form a quantum dot separated from the bulk by a tunneling barrier.  We assume that the lower two Landau levels are not pinched off and therefore do not backscatter at the point contact; the barrier region is assumed to have $\nu=2$, effectively forming a vacuum for the RR or ${\overline{\rm RR}}$ state.  For a finite system (the quantum dot) the edge modes acquire a discrete spectrum.\cite{Fiete:prl07,Ilan:prl08}  We focus on fluctuations of the quantum dot charge $Q \equiv e\langle \hat N_e \rangle$ near a degeneracy point in the energy: $E({N_e},S,B)=E({N_e}+1,S,B)$ where $N_e$ is the number of electrons on the quantum dot. The energy $E$ depends on dot area $S$, which may be altered by a gate potential shown in Fig.~\ref{fig:schematic}, and on the magnetic field $B$.  Adjusting either $S$ or $B$ may be used to achieve the desired degeneracy and also to slightly tune away from it. The charge of the dot can be measured capacitively. \cite{Ashoori:prl92,Berman:prl99,Duncan:apl99,DiCarlo:prl04}
 
We assume energy and temperature scales are much less than the level-spacing
of the dot edge states.  Then only two levels on the quantum dot (the degenerate or nearly degenerate ones) are important. These two levels act as an effective, local spin-1/2 degree of freedom.  The mapping of charge occupation on a quantum dot to an effective spin degree of freedom is not new\cite{Furusaki:prl02,Matveev:jetp90,Matveev:prb95,Kim:jpcm03,Oreg:prl03,Borda09,Klein_factors} and a number of related studies of quantum dots coupled to interacting one dimensional systems have been carried out.\cite{Kakashvili:prl03,Sade:prb05,Wachter:prb07,Lerner:prl08,Goldstein08,Goldstein09}  However, the physics resulting from coupling such a quantum dot to a RR or ${\overline {\rm RR}}$ state has only recently been touched upon in earlier work by the present authors.\cite{Fiete:prl08}  

For the remainder of this section we will study the Hamiltonian
\begin{equation}
\label{eq:H_full}
H=H_{\rm edge}+H_{\rm dot}+H_{\rm tun},
\end{equation}
where $H_{\rm edge}$ is the edge Hamiltonian of a RR or ${\overline {\rm RR}}$  state (including the $k=2$ MR and $\overline{\rm MR}$ states).  Here and henceforth, we ignore the $2$ filled lower Landau levels.  This is justified by the sequence of modes pinched off in a point contact.\cite{Dolev:nat08,Miller:nap07}  The dot Hamiltonian is
\begin{equation}
H_{\rm dot}=\epsilon_d\, d^\dagger d,
\end{equation}
where $d\, (d^\dagger)$ is the fermionic annihilation (creation) operator for this state. The level energy $\epsilon_d=0$ at the degeneracy point and $\epsilon_d \neq 0$ when one is tuned away from degeneracy. (Since we are focused on a level degeneracy point, the standard Coulomb charging term of the form $\frac{E_C}{2}(\hat N-N_e)^2$ does not play a key role in the physics of interest to us, although such a term is implicit in our analysis.)  The final term in \eqref{eq:H_full}, $H_{\rm tun}$, describes the tunneling of {\em electrons} between the edge of the bulk quantum Hall state and the quantum dot.  Its specific form depends on the quantum Hall state of the bulk, and it will determine the properties of the emergent Kondo model.  We will consider a number of important cases below. 

\subsection{Tunneling from the MR Pfaffian state}
The edge theory for the Pfaffian state
is the sum of a free, charged chiral bosonic sector
and a neutral Majorana (the `$\mathbb{Z}_2$ parafermion') sector. The edge Hamiltonian takes the form
\begin{equation}
\label{eq:edge_MR}
H_{\rm MR\, edge}=  \int dx  \:\left({v_c}\frac{2}{4\pi}  (\partial_x \phi(x))^2  \: + \: 
i v_n   \psi \partial_x \psi \right),
\end{equation}
where $v_n<v_c$ generally holds, $\phi$ is a real chiral boson,
and $\psi$ is a chiral Majorana fermion.  The Hamiltonian \eqref{eq:edge_MR} is equivalent to the sum of \eqref{eq:S_n} and \eqref{eq:S_c}.  The tunneling Hamiltonian for the MR state takes the form 
\begin{eqnarray}
H_{\rm tun}=t(d^\dagger {\Psi_{e,{\rm MR}}}(0) + {\Psi_{e,{\rm MR}}^\dagger}(0) d)\hspace{1cm}\nonumber \\
+ V {d^\dagger}d \,{\Psi^\dagger_{e,{\rm MR}}}(0){\Psi_{e,{\rm MR}}}(0),
\end{eqnarray}
where $t$ is the tunneling amplitude to the dot (which we have, without
loss of generality, assumed to be real). The tunneling is assumed to occur at the origin, $x=0$. The parameter $V$ represents the strength of the Coulomb repulsion between the edge and the dot,
and electron operator is given by Eq.\eqref{eq:Psi_MR}, $\Psi_{e,{\rm MR}}^\dagger=\psi e^{i2\phi}$.  It has scaling dimension 3/2.

As a result of the scaling dim of $\Psi_{e,{\rm MR}}$, $t$ is naively irrelevant for small $V$,
\begin{equation}
\label{eqn:RG-naive}
\frac{dt}{d\ell} = -\frac{1}{2}\,t + \frac{tV}{\pi{v_c}} + {\cal O}(t^3) .
\end{equation}
However, for $V$ sufficiently large, $t$ flows to the
$2$-channel Kondo fixed point, not to $t=0$. To see this, we we apply
to $H$ a unitary transformation similar to that of Emery and Kivelson
\begin{equation}
U=e^{2i {d^\dagger} d\, \phi(0)}.
\end{equation}
This rotates $\phi(0)$ out of the tunneling term.
$H$ now takes the form
\begin{multline}
\label{eqn:transformed-EK}
U H U^\dagger =
H_{\rm MR\, edge} +H_{\rm dot} +t\,\psi({d-d^\dagger} )\\
+\: \left(V-2{v_c}\right) {d^\dagger}d\, \partial_x \phi(0).
\end{multline}
For $V-2{v_c}=0$, this is a purely quadratic theory
which can be solved exactly. Thus, $t$ is clearly relevant
in this limit; it is actually relevant
over a range of values of $V$, as we discuss below.
Note that only the Majorana combination
${d-d^\dagger} $ couples to the the quantum Hall edge.
This is precisely the same feature which leads to
non-Fermi liquid behavior in the 2-channel
Kondo problem,\cite{Emery:prb92} as can be seen by direct comparison with \eqref{eq:H_K_final} for zero field, $h=0$, and channel {\em isotropy}, $\lambda_{\perp -}=0$.  

Equation \eqref{eqn:transformed-EK} therefore establishes that the tunneling of electrons from a MR state to a quantum dot is described by the channel {\em symmetric} 2-channel Kondo model.  Having the EK solution in hand, it is evident that the channel isotropy in \eqref{eqn:transformed-EK} follows from the unique Majorana fermion on the edge of the MR state, $\psi$, that appears in the electron operator. The absence of another edge Majorana to couple to the ``$d+d^\dagger$" combination follows directly from the topological properties of the MR Pfaffian state.  Therefore, the channel symmetric nature of the emergent 2-channel Kondo model is {\em topologically protected} by the MR state.  Typically, the channel symmetric limit requires delicate fine tuning and is very difficult to realize in experiment,\cite{Potok:nat07} but the topological protection of our situation is a boon.  Most importantly, though, the non-Fermi liquid physics of the emergent 2-channel Kondo model leads,
in principle, to a way to {\em identify} the fractional quantum Hall state as the MR Pfaffian state.

Another way to see the connection to the 2-channel Kondo model is to reverse the mapping of EK.  To this end it is useful to represent the two-level system on
the dot by a spin:$S^\dagger =\eta d^\dagger, S^-=d\eta$,
and $S^z=d^\dagger d -1/2$, where $\eta$ are Klein factors that ensure the proper commutation relations are achieved.\cite{Klein_factors}  The $\eta$ have the property that $\eta^\dagger=\eta$, $\eta^2=1$ and they anti-commute with fermions, {\it i.e}. their properties are like non-dynamical Majorana fermions. We apply a unitary transformation $U=e^{i\alpha S^z \phi(0)}$
to $H$ as before, but take $\alpha=2-\sqrt{2}$, to partially rotate $\phi(0)$ in the tunneling term,
giving
\begin{multline}
\label{eqn:transformed_eta}
U H U^\dagger =
H_{\rm MR\, edge} + \epsilon_d S^z + \left(V-
{v_c}\alpha \right) S^z \partial_x \phi(0)\\
+t(\eta\psi e^{-i \sqrt{2}\phi(0)}S^\dagger+
\psi \eta e^{i \sqrt{2}\phi(0)}{S^-}).
\end{multline}
When the term proportional to $t$ in Eq.~\eqref{eqn:transformed_eta} is computed to any order in perturbation theory, only even powers of $S^-S^\dagger$ and $S^\dagger S^-$ will appear and the Klein factor $\eta$ will disappear due to the relation $\eta^2=1$.  We may therefore safely drop it altogether, having already served its role in transforming fermionic operators $d$ to spin operators $S$,
\begin{multline}
\label{eqn:transformed}
U H U^\dagger =
H_{\rm MR\, edge} + \epsilon_d S^z + \left(V-
{v_c}\alpha \right) S^z \partial_x \phi(0)\\
+t(\psi e^{-i \sqrt{2}\phi(0)}S^\dagger+
\psi e^{i \sqrt{2}\phi(0)}{S^-}).
\end{multline}

We now compare this to the channel symmetric Kondo model:
\begin{equation}
\label{eqn:Kondo-coupling}
H_{\rm imp}= \lambda_\perp ( {J^+}(0)S^{-} + {J^-}(0)S^{+} )
+ \lambda_z {J^z}(0)S^{z} + h {S^z},
\end{equation}
where ${\vec S}$ is the impurity spin; ${\vec J}(0)$ is 
conduction electron spin density at the impurity site;
$\lambda_\perp$, $\lambda_z$ are the exchange
couplings, which are not assumed to be equal;
and $h$ is the magnetic field. The currents ${J^a}$ can be expressed in terms of a Majorana
fermion, $\psi$, and a free boson $\phi$,
\begin{equation}
\label{eqn:current-rep}
J^+=\sqrt{2}\psi e^{i \sqrt{2} \phi},{\hskip 0.2 cm}
J^-=\sqrt{2}\psi e^{-i \sqrt{2} \phi},{\hskip 0.2 cm}
J^z=\sqrt{2} \partial_x \phi,
\end{equation}
which is a special case of (\ref{ParafCurrents}).
The factors of $\sqrt{2}$ are present in this expression
because the boson $\phi$ is normalized here according to
Eq. \eqref{eq:edge_MR} so that $\text{dim}[e^{i\phi}]=1/4$,
rather than $\text{dim}[e^{i\phi}]=1/k=1/2$, as assumed
in Eq. \eqref{ParafCurrents}.

Substituting (\ref{eqn:current-rep}) into (\ref{eqn:Kondo-coupling}),
we see that our problem (\ref{eqn:transformed})
maps onto the $2$-channel Kondo model
with anisotropic exchange if we identify ${\lambda_\perp}=t$,
$\sqrt 2{\lambda_z}=V-(2-\sqrt{2}){v_c}$, and $h=\epsilon_d$.
For ${\lambda_z}<0$,
the Kondo model is ferromagnetic. In the ferromagnetic
Kondo model, the coupling to the impurity is irrelevant, as we naively
expected above (\ref{eqn:RG-naive}). However, when $V$
is sufficiently large, ${\lambda_z}>0$,
corresponding to the antiferromagnetic Kondo model.
In section \ref{sec:discussion}, we discuss the regime
$V-(2-\sqrt{2}){v_c}>0$ in terms of realistic parameters
for experiments.
In this case, the Hamiltonian is controlled in the infrared
by the exchange and channel symmetric antiferromagnetic
spin-1/2 $2$-channel Kondo fixed point.\cite{Affleck:prb92}
The non-Fermi liquid behaviors of the spin
susceptibility and magnetic field dependence of the zero-temperature magnetization
translate to the charge susceptibility and charge of the
quantum dot: \cite{Affleck:npb91}
\begin{eqnarray}
\chi_\text{charge} &\propto& \ln\left({T_K}/T\right)\: , \cr
{\Delta Q} &\propto& {V_G} \ln\left({k_B}{T_K}/{e^*}{V_G}\right).
\end{eqnarray}
Here, $\Delta Q = Q-e({N_e}+\frac{1}{2})$ is the charge on
the dot measured relative to the average electron number
at the degeneracy point of the energy.
The Kondo temperature $T_K$ is the energy scale at which
the coupling becomes ${\cal O}$(1). In the exchange isotropic Kondo problem, in which
the interaction $\lambda$ is marginal at tree level,
${T_K}\propto\exp(-\pi{v_n}/\lambda)$.
In a Majorana fermion resonant level model, in which the tunneling
$\lambda$ has scaling dimension $1/2$, ${T_K}\propto\lambda^2$.
In our problem, however, due to the RG equation
(\ref{eqn:RG-naive}),
together with $dV/d\ell = {t^2}/\pi{v_c}$,
we have a Kosterlitz-Thouless-like set of RG equations.
For $V\gg \pi{v_c}/2$, the flows in the $V-t$ plane are
nearly vertical, so that 
${T_K}\propto t^{1/\left(\frac{V}{\pi{v_c}} - \frac{1}{2}\right)}$.
However, closer to the Kosterlitz-Thouless point, we expect the flow to
strong coupling to be slower so that, for instance,
along the KT separatrix, an exponential
form such as ${T_K}\propto e^{-c/t}$ would hold for some
constant $c$.

\subsection{Tunneling from the RR state}

The analysis for electron tunneling between a RR state for
general $k$ and the quantum dot is almost identical to that for $k=2$.  The edge theory for filling fraction $\nu=k/(k+2)$ is
\begin{equation}
\label{eq:edge}
H_{\rm edge}=  \frac{(k+2)/k}{4\pi} v_c\int dx  \:  (\partial_x \phi(x))^2  \: + \: 
H_{\mathbb{Z}_k},
\end{equation}
where $H_{\mathbb{Z}_k}$ describes the (interacting, except for $k=2$)
neutral $\mathbb{Z}_k$ parafermionic sector of the edge.
The tunneling term takes the form
\begin{eqnarray}
H_{\rm tun}=t(d^\dagger {\Psi_{e,{\rm RR}}}(0) + {\Psi_{e,{\rm RR}}^\dagger}(0) d)\hspace{1cm}\nonumber \\
+ V {d^\dagger}d \,{\Psi^\dagger_{e,{\rm RR}}}(0){\Psi_{e,{\rm RR}}}(0),
\end{eqnarray}
and $H_{\rm dot}$ is the same as before.  The electron operator $\Psi_e^\dagger=\psi_1 e^{i\frac{k+2}{k}\phi}$, as given in \eqref{eq:Psi_RR}, and has scaling dimension 3/2.  The crucial difference between the RR state and the MR state is that the $\mathbb{Z}_k$ parafermion $\psi_1$ has replaced the $\mathbb{Z}_2$ Majorana fermion.

As before, we apply a unitary transformation
$U=e^{i\alpha S^z \phi(0)}$ to $H$, which now takes the form,
\begin{multline}
\label{eqn:transformed-RR}
U H U^\dagger =
H_{\rm edge} +H_{\rm dot} + \left(V-
{v_c}\alpha \right) S^z \partial_x \phi(0)\\
+t(\psi_1^\dagger e^{-i\tilde \alpha\phi(0)}S^\dagger+
\psi_1e^{i\tilde \alpha\phi(0)}{S^-}),
\end{multline}
where $\tilde \alpha\equiv \frac{k+2}{k}-\alpha$.
The choice $\alpha^* = \frac{k+2}{k}-\sqrt{\frac{k+2}{k}}\sqrt{\frac{2}{k}}$
makes the connection to the $k$-channel Kondo problem explicit
because the SU(2)$_k$ current operators
can be represented in terms of the $\mathbb{Z}_k$ parafermions:
\begin{equation}
J^+=\sqrt{k}\psi_1 e^{i \beta \phi},\;J^-=\sqrt{k}\psi_1^\dagger e^{-i \beta \phi},\;
J^z={k \over 2}\beta \partial_x \phi,
\end{equation}
where $\beta=\sqrt{2(k+2)}/k$. [As in the MR case,
the boson $\phi$ is normalized differently here
than in Eq.~\eqref{ParafCurrents}: according to
Eq.~\eqref{eq:edge}, $\text{dim}[e^{i\phi}]=k/2(k+2)$ 
rather than $\text{dim}[e^{i\phi}]=1/k=1/2$, as assumed
in Eq.~\eqref{ParafCurrents}.]
Substituting these expressions into 
(\ref{eqn:Kondo-coupling}), we see that our problem
(\ref{eqn:transformed-RR})
is equivalent to the channel {\em symmetric} $k$-channel Kondo problem
if we identify ${\lambda_\perp}=t$,
$\beta {\lambda_z}=V-{v_c}\alpha^* $, and $h={\epsilon_d}$.
For $V>{v_c}\alpha^* $, this is the antiferromagnetic Kondo
problem, which has an intermediate coupling fixed point.
Thus, we see that the Read-Rezayi states offer a novel scenario to
realize the non-Fermi liquid behavior of the $k$-channel Kondo model,
\begin{eqnarray}
\chi_\text{charge} &\propto& T^{-(k-2)/(k+2)},\cr
{\Delta Q} &\propto& {V_G}^{2/k},
\end{eqnarray}
which would otherwise require an incredible amount of fine-tuning
for $k\geq 3$. Moreover, observing the predicted non-Fermi liquid behavior
would provide strong evidence that the quantum Hall state
is of the Read-Rezayi type.  Again, we emphasize that the non-Fermi liquid physics of the channel symmetric $k$-channel Kondo model is {\em topologically protected} by the RR state:  Its edge theory has a unique parafermion in the electron operator.

As we will now see, the non-Fermi liquid physics is no longer topologically protected when one considers the particle-hole conjugate states.
The basic reason is that the particle-hole conjugate states have
multiple electron operators which need not have the same
tunneling amplitudes.

\subsection{Tunneling from the ${\overline {\bf MR}}$ anti-Pfaffian state}

The edge theory of the ${\overline {\rm MR}}$ anti-Pfaffian state is: \cite{Levin:prl07,Lee:prl07} 
\begin{equation}
{\cal L}_{\overline{\rm MR}}
= \frac{2}{4 \pi} {\partial_x}{\phi}
(i{\partial_\tau}+v_c{\partial_x}){\phi}
+\sum_{a=1}^3 i{\psi_a}(-i{\partial_\tau}+
v_n{\partial_x}){\psi_a},
\end{equation}
as discussed in Sec.~\ref{sec:edge_theory}. There are three different dimension-$3/2$
electron operators, ${\psi_a} e^{2i\phi}$ for $a=1,2,3$. The tunneling Hamiltonian is,
\begin{equation}
H_{\rm tun}= \sum_{a=1}^3\left(
{t_a}\psi_a e^{-2i\phi} d^\dagger + \text{h.c.}\right)
+ V {d^\dagger} d {\partial_x} {\phi}.
\end{equation}
Although the edge theory has an emergent SU(2) symmetry,
and the three electron operators ${\psi_a}$ form a triplet
under this symmetry, this symmetry is not fundamental
and the tunneling operators do not have to respect it.

Performing a unitary transformation, as before, to rotate out
the $\phi$ dependence of the first term, we obtain
$U H {U^\dagger} = H_\text{edge} + H_\text{dot} +
{\tilde H}_\text{tun}$ where 
\begin{multline}
\label{eqn:anti-Pfaff-Toulouse}
{\tilde H}_\text{tun} =
\sum_{a=1}^3\left({t_a}\psi_a d^\dagger + \text{h.c.}\right)
+ \left(V-2{v_c}\right) {d^\dagger} d {\partial_x} {\phi},\\
= i{\chi^{}_1}\left( {\lambda_1} ({d^\dagger}-d)/i  + {\lambda'_1}({d^\dagger}+d)\right)\\
+ i{\lambda_2}{\chi^{}_2}({d^\dagger}+d)
\:+ \left(V-2{v_c}\right) {d^\dagger} d \, {\partial_x} {\phi},
\end{multline}
where ${\chi_1}={u_a}{\psi_a}/\sqrt{u^2}$, ${\chi_2}={w_a}{\psi_a}/\sqrt{w^2}$,
${u_a}=\text{Re}\,{t_a}$, ${v_a}=\text{Im}\,{t_a}$, ${w_a}={v_a}-{u_a}(u\cdot v/{u^2})$,
${\lambda_1}=\sqrt{u^2}$, ${\lambda_2}=\sqrt{w^2}$, ${\lambda'_1}=u\cdot v/\sqrt{u^2}$,
and $\{{\chi_1},{\chi_2}\}=0$.
Note that, for generic $t_a$, both $a=({d^\dagger}+d)/\sqrt{2}$
and $b=({d^\dagger}-d)/i\sqrt{2}$
couple to the edge modes,
as in the {\it one-channel} Kondo model.
Thus, we expect that (\ref{eqn:anti-Pfaff-Toulouse})
is also controlled by the one-channel Kondo fixed point.

This is in contrast to the Pfaffian case, in which only
$b$ couples to the edge modes, as in the channel symmetric two-channel Kondo model. One might naively expect that both
$a$ and $b$ would couple to the edge modes even in the Pfaffian case
if the tunneling amplitude is $t$ is not purely real. However,
for $t=|t|e^{i\theta}$
we could always find a linear combination
$a\cos\theta + b\sin\theta$ which does not couple.
The problem in the anti-Pfaffian case is that
the linear combination $a\cos{\theta_1} + b\sin{\theta_1}$
which does not couple to $\psi_1$ will, in general, couple
to $\psi_2$ and $\psi_3$. Only in the special case in which
all three tunneling amplitudes have the same phase will
two-channel Kondo behavior result.

The charge susceptibility and charge of the
quantum dot have the temperature and voltage dependence
characteristic of a Fermi liquid:
\begin{equation}
\chi_\text{charge} \propto \text{const.}\:\:, \hskip 0.5 cm
{\Delta Q} \propto {V_G}.
\end{equation}
Consequently, measurements of the behavior of the dot would
distinguish the Pfaffian and anti-Pfaffian states.

\subsection{Tunneling from the ${\overline {\bf RR}}$ state}

The edge theory of the ${\overline {\rm RR}}$ state is given by the sum of \eqref{eq:S_L} with $\nu=2/(k+2)$ and \eqref{eq:S_n_k}.  The tunneling Hamiltonian takes the form
\begin{equation}
H_{\rm tun}= \sum_{m=-k/2}^{k/2}\left(
{t_m}\chi_{j=k/2}^m e^{i\frac{k+2}{2}\phi} d + \text{h.c.}\right)
+ V {d^\dagger} d {\partial_x} {\phi}.
\end{equation}
It is useful to rewrite this expression using the parafermion
representation of SU(2)$_k$ (e.g. using
(\ref{ParafCurrents})). Then
$\chi_{j=k/2}^m = \psi^{}_{m+k/2}\, e^{-i m {\phi_\sigma}/2}$
where $\psi^{}_{0}=\psi^{}_{k}\equiv 1$ and $\psi^{}_{1}, \psi^{}_{2},
\ldots, \psi^{}_{k-1}$ are the parafermion operators. The tunneling Hamiltonian then takes the form
\begin{multline}
H_{\rm tun} = \sum_{m=-k/2}^{k/2}\left(
t^{}_{m}\psi^{}_{m+k/2}\, e^{-i m {\phi_\sigma}/2}
e^{i\frac{k+2}{2}\phi} d + \text{h.c.}\right)\\
+ \,\,V {d^\dagger} d {\partial_x} {\phi}\,.
\end{multline}
Separating the $m=1-k/2$ term from the sum,
we have,
\begin{multline}
H_{\rm tun}
= t^{}_{1-k/2}\,\psi^{}_{1}\, e^{i(k-2) {\phi_\sigma}/2}
e^{i\frac{k+2}{2}\phi} d + \text{h.c.} \,+\,\,
V {d^\dagger} d {\partial_x} {\phi}\\
\sideset{}{'}\sum_{m=-k/2}^{k/2} \left(
t^{}_{m}\psi_{m+k/2} e^{-i m {\phi_\sigma}/2}
e^{i\frac{k+2}{2}\phi} d + \text{h.c.}\right)\,.
\end{multline}
where the prime on the summation indicates that the sum does not
include the $m=1-k/2$ term.  Performing a unitary transformation,
we can rewrite this as:
\begin{multline}
\label{eqn:anti-k-res-model}
H_{\rm tun}
= t^{}_{1-k/2}\,\psi^{}_{1}\, d + \text{h.c.} \,+\,\,
\left(V-(1+k/2){v_c}\right) {d^\dagger} d\, {\partial_x} {\phi}\\
+ (1-k/2){v_n} {d^\dagger} d \, {\partial_x} {\phi_\sigma}\,+\\
\sideset{}{'}\sum_{m=-k/2}^{k/2} \left(
t^{}_{m}\psi_{m+k/2} e^{-i (m-1+k/2) {\phi_\sigma}/2}
e^{i\frac{k+2}{2}\phi} d + \text{h.c.}\right).
\end{multline}
The first line of (\ref{eqn:anti-k-res-model}) is essentially
the channel symmetric $k$-channel Kondo model, after unitary
transformation, as in (\ref{eqn:transformed-RR}) with
$\tilde{\alpha}=\frac{k+2}{2}$. The second and third lines
may be viewed as perturbations of this model. There are
$k$ couplings $t_m$, $m\neq 1-k/2$, and it is tempting to
identify them with the possible channel anisotropies in the
$k$-channel Kondo problem. However, we do not have
a direct mapping and, indeed, we do not expect a simple
mapping of this form since channel asymmetry in the Kondo
model would necessarily involve operators in the SU(k)$_2$
flavor sector, which does not enter our model. However,
the third line of (\ref{eqn:anti-k-res-model}) is certainly a relevant
perturbation, so it will drive the system away from the
channel symmetric $k$-channel fixed point. It seems likely
that it will drive the system to the one-channel fixed point,
i.e. that there will be no boundary entropy left, since that is
the most stable fixed point and the $k$ relevant operators in
Eq.~\eqref{eqn:anti-k-res-model} would be expected to destabilize
any other possible fixed point. However, whether or not the
one-channel Kondo behavior governs the ultimate low-energy
fixed point, tunneling to the ${\overline {\rm RR}}$ state will
not be governed by the channel symmetric $k$-channel
Kondo model.

\subsection{Other candidate states at $\nu=5/2$ and $\nu=12/5$}
Before leaving the problem of a quantum dot coupled to a {\em single} edge of a  fractional quantum Hall state, we briefly return to filling fraction $\nu=5/2$.  Besides the Moore-Read Pfaffian and the anti-Pfaffian, there are at least three other candidate states for $\nu=5/2$, the non-Abelian SU(2)$_2$ NAF state,\cite{Wen:prl91,Blok:npb92} the Abelian K=8 strong coupling state,\cite{Halperin83} and the Abelian Halperin (3,3,1) state.\cite{Halperin83}  While numerically, the Pfaffian and anti-Pfaffian appear to be favored,\cite{Morf98,Rezayi00,Wang09,Bonderson09,Feiguin:prb09,Wojs:prb09} it is useful discuss what type of behavior we would expect in our set-up for these other states.  In a recent work by Bishara {\it et al.}  the signatures in quantum Hall interferometry for each of these 5 candidate states were discussed.\cite{Bishara09}

In order to anticipate what behavior to expect for these other candidate states, it is useful to recall the feature that led to the stable 2-channel Kondo behavior for the Pfaffian:  a unique edge electron operator that was built in part from the Majorana fermion.  Since the electron operator and the SU(2) currents only differ by the scaling dimension of the bosonic charge sector portion of the electron operator (which can be changed with a unitary transformation in the tunneling Hamiltonian as we saw earlier), the Pfaffian is guaranteed to map onto the channel isotropic version of the 2-channel Kondo model.  When the electron operator was no longer unique and/or was not built from a parafermion (as is the case for the $\overline{\rm MR}$ or $\overline{\rm RR}$ states) then the system generically flows to a single-channel Kondo model.  We expect the single-channel Kondo fixed point to be the ultimate fate of the non-Abelian SU(2)$_2$ NAF state, the Abelian K=8 strong coupling state, and the Abelian Halperin (3,3,1) state.
This is true, even though in states like the (3,3,1) state where the spin degree of freedom is active and one would expect a 2-channel Kondo model to be realized (because the situation is similar to that discussed in Ref.~[\onlinecite{Matveev:jetp90,Matveev:prb95}]).  However, this 2-channel Kondo model will not be ``topologically protected" in the way that it is for the Pfaffian.  Because of residual Zeeman coupling to the electron's spin in the quantum Hall state the spin up and spin down edge modes will be slightly shifted with respect to one another on the edge.  This will lead to different tunneling matrix elements between the edge and dot for different spin orientations, breaking the channel symmetry in the effective 2-channel Kondo model.  Thus, the low energy fixed point will be described by a single-channel Kondo model, rather than the 2-channel version. In the case of the
non-Abelian SU(2)$_2$ NAF state, the analysis is
nearly identical to the analysis of the anti-Pfaffian state.
The edge theories of the two states differ only in the chiralities
of the neutral modes, and physics at a single point contact
is completely blind to the chirality of the edge modes.
Thus, single channel Kondo behavior is obtained.
The edge theory of the Abelian K=8 strong coupling state is
a single chiral boson, which maps to the single-channel
Kondo problem (after Toulouse transformation) if the
edge-dot repulsive interaction is sufficiently strong
(and otherwise flows to the ferromagnetic Kondo fixed point
at which the edge and dot decouple, as in all of the
models which we consider here).

Finally, the state at $\nu=12/5$ has another non-Abelian candidate (besides the $\overline{\rm RR}$ state at k=3), a Bonderson and Slingerland (BS) state in the hierarchy built on the Pfaffian.  Due to the presence of other edge modes in the BS state at  $\nu=12/5$ (relative to the Pfaffian) we expect the low-energy fixed point to again be described by the single-channel Kondo model.  While both the $\overline{\rm RR}$  and BS state are expected to have the same low-energy fixed point, the crossover at higher energy scales should be governed by the 3-channel Kondo model for $\overline{\rm RR}$ and the 2-channel Kondo model for BS.   However, due to the unknown intrinsic exchange anisotropies involved, it is likely to be very difficult to conclude much in experiment from this intermediate energy behavior.  

\subsection{Tunneling through a Quantum Dot}

\begin{figure}[ht]
\includegraphics[width=.85\linewidth,clip=]{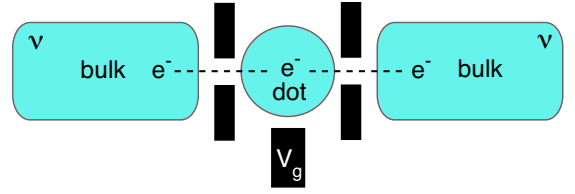}
 \caption{\label{fig:transport} (color online) Schematic of our model for electron tunneling through a quantum dot.  Gates are shown in black. They may be used to form a point contact to pinch off the dot from the rest of the quantum Hall bulk.  The gate on the bottom of the figure may be used to shift the energy levels of the dot by changing its area $S$ through a change in the gate voltage $V_g$.  The bulk is assumed to be at filling fraction $\nu=2+k/(k+2)$ or $\nu=2+2/(k+2)$. The white region between the dot
and the bulk is assumed to be at $\nu=2$.}
 \end{figure}

We now consider the situation of bulk $\nu=2+k/(k+2)$
quantum Hall states on either side of a quantum dot, as shown in Fig.\ref{fig:transport}. Such a configuration
will arise in a two point contact interferometer if the
backscattering at the two point contacts is increased until
they are both near pinch-off (at $\nu=5/2$, this will be pinch-off
of the $\nu=1/2$ state in the second Landau level, while the
filled lowest Landau levels are not pinched off).
The magnetic field or a side-gate voltage must then be tuned
to a degeneracy point of the dot which has been created
between the two bulk states. 

For the sake of concreteness,
we first consider the MR Pfaffian state. The edge Hamiltonian is:
\begin{equation}
H_{\rm MR\, edges}=  \sum_{i=1,2}\int dx  \:\left({v_c}\frac{2}{4\pi} 
(\partial_x {\phi_i}(x))^2  \: + \: 
i v_n   {\psi_i} \partial_x {\psi_i} \right),
\end{equation}
where $i=1,2$ are the two bulk states to the left and right of the dot.
The tunneling Hamiltonian takes the form 
\begin{multline}
H_{\rm tun}=\sum_{i=1,2}{t_i}(d^\dagger {\psi_i}(0)e^{2i{\phi_i}(0)}
+  {\psi_i}(0)e^{-2i{\phi_i}(0)} d) \\
+ {V_i} {d^\dagger}d \,{\partial_x}{\phi_i}(0).
\end{multline}
This model is tantalizingly close to the $4$-channel Kondo
problem, but it is not quite equivalent to it. If we were to consider
the bosonic analogue of this problem -- bosons as $\nu=1$ --
then the Hamiltonian would instead be
\begin{multline}
H_{\rm bosonic\,MR} =  \sum_{i=1,2}\int dx  \:\left({v_c}\frac{1}{4\pi} 
(\partial_x {\phi_i}(x))^2  \: + \: 
i v_n   {\psi_i} \partial_x {\psi_i} \right)\\
+ \sum_{i=1,2}{t_i}(d^\dagger {\psi_i}(0)e^{i{\phi_i}(0)}
+  {\psi_i}(0)e^{-i{\phi_i}(0)} d) \\
+ {V_i} {d^\dagger}d \,{\partial_x}{\phi_i}(0).\\
\end{multline}
The second and third lines could then be written as,
\begin{equation}
 \sum_{i=1,2}{t_i}(S_+ {J_-^i}
+  S_-  {J_+^i})
+ {V_i} S_z J_z^i,
\end{equation}
where $J_a^1$ and $J_a^2$ both generate SU(2)$_2$
Kac-Moody algebras. For ${t_1}={t_2}$, ${V_1}={V_2}$,
this is just a cumbersome way of writing the $4$-channel Kondo
Hamltonian. However, in the $\nu=1/2$ fermionic version
of this problem, which is relevant to us, we need to
perform a unitary transformation generated by
\begin{equation}
\label{eqn:transport-unitary}
U=e^{(2-\sqrt{2})i {d^\dagger} d\, {\phi_1}(0)},
\end{equation}
in order to transform the $t_1$ term to Kondo form:
\begin{multline}
\label{eqn:asymmetric-transport}
U H U^\dagger =
{t_1}(d^\dagger {\psi_1}(0)e^{\sqrt{2}i{\phi_1}(0)}
+  {\psi_1}(0)e^{-\sqrt{2}i{\phi_1}(0)} d)\\
+ {t_2} (d^\dagger {\psi_2}(0)e^{2i{\phi_2}(0)}
e^{-(2-\sqrt{2})i  {\phi_1}(0)}\\
+  {\psi_2}(0)e^{-2i{\phi_2}(0)}
e^{(2-\sqrt{2})i  {\phi_1}(0)} d).
\end{multline}
This unitary transformation makes the $t_2$ term complicated
(and highly irrelevant, at least naively). Conversely, we could
bring the $t_2$ term into two-channel Kondo form, at the cost
of making the $t_2$ term, at the cost of making the $t_1$
term complicated. In the case of more general RR states, the same situation is appears: we have representations in which
the coupling to either bulk state of these terms is simple and
of $k$-channel Kondo form, but we don't have a representation
in which both couplings are simple and tractable. It is thus clear that the situation in Fig. \ref{fig:transport} does not immediately map to a $2k$-channel Kondo model.  The nature of its fixed point(s) is
an interesting open problem.

In the case of ${t_1}\neq{t_2}$, the low temperature physics can exhibit some interesting crossovers. Suppose that ${t_1}\gg {t_2}$, but both
are still small (compared to the charging energy of the dot and
all other microscopic scales). Then it makes sense to perform
the unitary transformation (\ref{eqn:transport-unitary}) in order
to bring the Hamiltonian to the form (\ref{eqn:asymmetric-transport}).
In this form, $t_1$ is relevant and flows to the $2$-channel Kondo
fixed point, while $t_2$ is irrelevant and flows to zero.
The most salient feature in this limit will be that the
dot does not, as the temperature is lowered, decouple from
both bulk QH states but, instead, from only one of them.
Consequently, Coulomb blockade peaks, rather than
narrowing as the temperature is lowered, may broaden
instead since there will not be a completely isolated dot
at $T=0$.

\section{Discussion}
\label{sec:discussion}

The intermediate coupling fixed points which we found in this paper may seem somewhat mysterious, so we suggest a physical picture which may help explain them. In order to do so, we draw on the ideas of Ref. [\onlinecite{Fendley09}], in which it was shown that quantum
Hall edge states and their interaction with bulk quasiparticles could be understood in terms of boundary conformal field theory as follows. If the boundary of a quantum Hall droplet is treated as a 1-d system by `squashing' the edge down and temporarily ignoring the 2-d bulk,
then the 1-d system lives on an interval and, therefore, has conformally-invariant boundary conditions at the two ends of the interval (since there was clearly no length scale
introduced in the process of reduction to a 1-d system). (Here, `droplet' refers to the bulk quantum Hall state, not the `dot'.)   Different possible conformally-invariant boundary conditions
correspond to different possible quasiparticles in the 2-d bulk, which is one way in which the system evinces, through bulk-edge interaction, the fact that it is not really 1-d. In Abelian states,
this is relatively trivial and can usually be ignored, but in non-Abelian
states, this effect can be significant. In particular, different possible conformally-invariant
boundary conditions have different boundary entropies, which are given by the quantum dimensions of the associated bulk quasiparticles. Returning now to our new fixed points, we note that the ground state entropy change associated with our new fixed points
is precisely equal to the entropy drop expected if the two-fold degeneracy
of the dot is lifted and the boundary entropy of the edge excitations of the droplet
{\it increases} by $\ln d$, where $d=2\cos\pi/(k+2)$ is the quantum dimension of the minimal charge quasiparticle. (Here, we refer to the boundary entropy at the point at which tunneling to the dot occurs when the edge is `squashed' to a 1-d interval.)  We emphasize that the change in boundary entropy is measured via a change in boundary conditions of the {\em droplet}/bulk quantum Hall state edge that provides the electrons tunneling to the dot, and so is {\em independent of the details} of the electronic structure on the dot.

This suggests the following picture for our fixed point, which relies on the fact that the fixed point occurs at intermediate dot-bulk edge coupling so that the charge on the dot fluctuates and is not necessarily integral.  (Again, we note this is true independent of the electronic structure of the dot, provided that it is a degenerate two-level system in the absence of tunneling.  In particular, it does not matter whether it is in a quantum Hall state or not.)  Thus, we can imagine that a quasiparticle-quasihole pair is nucleated so that the quasiparticle is on the droplet and the quasihole is on the dot (or vice versa).  This is true, even though at ``high energies" it is electrons that are tunneling, not quasi-particles.  The presence of the quasiparticle on the droplet changes the
boundary entropy by according to the arguments of Ref. [\onlinecite{Nayak08,Fendley09}].
The presence of the quasihole on the dot lifts the degeneracy between the $N$ and $N+1$-electron states, thereby changing the boundary entropy by $-\ln 2$.  To see this, consider, for the sake of concreteness, a Pfaffian state. The energy on a small Pfaffian dot can be modeled by
$E(N)=E_C{(n-{n_0})^2}/2 + {v_n}n/2R$ where $n=0,1$
corresponds to $N$ or $N+1$ electrons, $R$ is the length of the edge,
and $E_C$ is the Coulomb charging energy. The first term
is the Coulomb energy, with an offset $0<{n_0}<1$
while the second term is the energy in the neutral sector
(associated with creating a Majorana fermion when an
electron is added). By tuning ${n_0}=(1+2\frac{v_n}{RE_C})/2$,
we can make the $N$ and $N+1$ electron states degenerate
in energy. However, when a charge $e/4$ quasiparticle
is added to the dot, the neutral energy vanishes due to
the existence of a dot edge zero mode. Consequently, the
degeneracy is lifted. The total entropy drop is thus $-\ln 2/d$,
as predicted for the multi-channel Kondo model.\cite{Affleck:npb91,Affleck_2:npb91}

We should emphasize that the precise nature of the electronic states on the quantum dot are not important for our basic result of multi-channel Kondo physics in the electron tunneling on and off the dot.  In particular, if there is coupling between the edge of the dot and the bulk of the dot, the physics will remain unchanged provided that the energy of other states (not involved in the 2-fold level degeneracy) is more than the tunnel broadening and $k_B T$ away in energy.  In this case, the level ``d" will correspond to some hybridization of edge and bulk states.  In fact, the fractional quantum Hall state in the partially filled Landau level in the dot can even be destroyed by finite size effects (from being too small for example) and it would not affect our conclusions, provided again that the nearby level spacing is larger than tunnel broadening and temperature.  However, there is one detail that is important to our result, and that is that the completely filled lower Landau levels pass freely under the point contact.  This detail is important because it means that the tunneling from the bulk quantum Hall edge to the quantum dot only occurs from the ``Read-Rezayi" part and not the filled lower Landau levels.  If the two integer quantum Hall edges are also backscattered at the point contact, they should also be included in the edge electron tunneling operators (note the plural!) to the quantum dot.  These additional electron tunneling processes will drive the system towards the single-channel Kondo model at all filling fractions since the ``channel isotropy" coming from the parafermion part of the electron operator will no longer protect the multi-channel Kondo physics when these additional channels are present. In effect, we expect the physics to be similar to the case of the anti-Pfaffian where the edge electron operator is not unique and the system is described by the single-channel Kondo model at the lowest energies.

There are at least two other scenarios (of which we are aware) where 2-channel Kondo physics emerges in the context of quantum Hall states: Fradkin and Sandler discuss a junction of a $\nu=1/3$  fractional quantum Hall state and a normal metal.\cite{Sandler:prb01}   Fendley, {\it et al.} discuss the IV characteristics of a point contact in a Pfaffian state and show that the crossover from weak to strong backscattering
of charge $e/4$ quasiparticles is described by a variant of the 2-channel Kondo model.\cite{Fendley:prl06,Fendley:prb07} 
In both cases, the physics is quite distinct from what we discuss in this work.
We are unaware of any other works considering multi-channel Kondo models derived from the general Read-Rezayi states, or other candidate non-Abelian fractional quantum Hall states. 

In order to see multi-channel Kondo physics in our setup for the MR state,
we need $V>(2-\sqrt{2})v_c$.  The following is a very crude estimate of the
relevant parameters.
We take ${v_c}\approx 10^5$m/s (although smaller values are possible
if the edge confining potential is smoother), which implies
$\hbar {v_c}\approx 10^{-10} \,\text{eV}\cdot\text{m}$. The coupling
$V$ is the Coulomb energy for an electron on the dot and charge
per unit length ${\partial_x}\phi_c$ on the edge in the vicinity of $x=0$. 
If $r$ is the distance between the dot and the bulk state,
and the magnetic length $\ell_0$ is the short-distance cutoff for
edge physics, then we have $V\propto \frac{e^2}{\epsilon r}\,\ell_0$.
For $r\sim\ell_0$, $V\approx 10^{-10}\,\text{eV}\cdot\text{m}$.
Thus, $V$ and $v_c$ are comparable, and by tuning the edge velocity
or the bulk-dot distance, it should be possible to arrange $V>(2-\sqrt{2})v_c$.
We emphasize that this estimate is extremely crude because it is difficult to accurately estimate the strength of the interaction $V$ as it depends on non-universal details of the edge, including screening effects from the gates and possibly the edge modes themselves. 
At any rate, by the same logic, it should be possible to make the exponent
$1/\left(\frac{V}{\pi{v_c}} - \frac{1}{2}\right)$ which appears in
the Kondo temperature,
${T_K}\propto t^{1/\left(\frac{V}{\pi{v_c}} - \frac{1}{2}\right)}$,
of order $1$. If, for instance, it is precisely equal to $1$, we will
have ${T_K}\propto t$. In edge tunneling experiments,
$t$ values are often in the range $t/{\ell_0}\sim 10 \,\text{K}$,
so we expect that the multi-channel Kondo fixed point will be observable for
accessible temperatures.

A complicating issue in real experiments is possible edge reconstruction, which is expected if the edges are sufficiently smooth.\cite{Chklovskii:prb92,Chamon:prb94,Wan:prl02,Joglekar:prb03,Orgad:prl08}  When edge reconstruction occurs, pairs of counter-propagating edge modes appear that do not affect the Hall conductance, but do affect the edge Hamiltonian.  The presence of these modes could in principle destabilize the multi-channel Kondo physics we have discussed here at the lowest energies, but one generally expects their coupling to the dot to be much weaker due to their greater spatial separation from it.  In that case, it may be that even if edge reconstruction is present it will not have much effect over a fairly large range of temperatures and the multi-channel Kondo physics will still be observable.

Finally, while the observation of non-Fermi liquid physics in the thermodynamics of the dot would provide strong evidence for a non-Abelian state of Moore-Read or Read-Rezayi type (at the appropriate filling fraction), the converse is not true:  lack of non-Fermi liquid properties could result from the presence of a particle-hole conjugate state, strong edge reconstruction, or even an Abelian quantum Hall state.  Thus, if Fermi liquid properties are observed, further studies must be done to determine if the state is non-Abelian.

\section{Summary}
\label{sec:summary}

In summary, we have shown that a quantum dot coupled via tunneling to
a Pfaffian quantum Hall state realizes the channel symmetric $2$-channel Kondo model
while a quantum dot coupled to a Read-Rezayi state of filling factor
$\nu=2+k/(k+2)$ leads to a channel symmetric $k$-channel Kondo problem, both without any fine tuning of parameters.  These systems will thus exhibit all the known non-Fermi liquid properties in their thermodynamics, expressed through the charge on the dot, which may be measured capacitively. Because the coupling of a quantum dot to an anti-Pfaffian
state generically exhibits Fermi liquid properties, our results may be used to distinguish between the two leading candidate states for $\nu=5/2$: the Pfaffian and the anti-Pfaffian.

Our central results can be understood within the context of the Emery-Kivelson (EK) solution to the 2-channel Kondo model.  In the EK analysis, a 2-channel Kondo model is mapped to a resonant level coupled to some gapless degrees of freedom.  In the channel symmetric case, the gapless mode is a Majorana fermion and non-Fermi liquid impurity physics results.  Thus, 2-channel Kondo physics is governed by a Majorana resonant level.  In the case of channel asymmetry, the level effectively couples to a Dirac fermion and Fermi liquid impurity physics is found, characteristic of the 1-channel Kondo model.  

In our work, we reverse the process and ask ``To what quantum impurity model does a resonant level coupled to a RR state correspond?".  For $k=2$ the RR state is the MR Pfaffian state, which has a {\em single} Majorana mode on its edge.  This Majorana mode turns out to play exactly the same role as the Majorana mode in the EK analysis.  For the RR states at general $k> 2$ the Majorana is replaced by a $\mathbb{Z}_k$ parafermion (which is a Majorana for $k=2$) which takes the place of the Majorana in the $k$-channel Kondo model.  The uniqueness of the parafermion mode in the edge of the RR state encodes the physics of channel isotropy in the effective multi-channel Kondo model.  As a result, the usually fragile multi-channel Kondo physics is  ``topologically protected" by the non-Abelian quantum Hall state.  For particle-hole conjugate states, the low-energy physics is governed by physics other than the channel symmetric multi-channel Kondo models.  Finally, we find that the transport through a quantum dot between two non-Abelian quantum Hall states is not governed by a $2k$-channel Kondo model, but rather by a fixed point that we could not determine with confidence.  Its properties are an interesting topic for future study.

\acknowledgments
We acknowledge helpful discussions with E. Ardonne, J. Eisenstein, L. Fidkowski,
A. Ludwig, and G. Refael.   GAF was supported by the Lee A. DuBridge Foundation and ARO grant W911NF-09-1-0527.
\vskip -0.2 cm


\end{document}